\UseRawInputEncoding
\documentclass[reprint,aps,superscriptaddress,prl]{revtex4-2}
\usepackage{graphicx}
\usepackage[colorlinks=true,pdfborder={0 0 0},linkcolor=blue]{hyperref}
\usepackage{url}
\begin{document}
\newcommand{\HY}[1]{{\color{blue}{ HY: #1}}}
\newcommand{\todo}[1]{{\color{red}{ #1}}}
\title{Quantum secret sharing in a triangular superconducting quantum network}

\author{Haoxiong Yan}
\altaffiliation[Present address: ]{Applied Materials, Inc, Santa Clara, CA 95051, USA}
\affiliation{Pritzker School of Molecular Engineering, University of Chicago, Chicago IL 60637, USA}

\author{Allen Zang}
\affiliation{Pritzker School of Molecular Engineering, University of Chicago, Chicago IL 60637, USA}

\author{Joel Grebel}
\altaffiliation[Present address: ]{Google Quantum AI, 301 Mentor Dr, Goleta, CA 93111, USA}
\affiliation{Pritzker School of Molecular Engineering, University of Chicago, Chicago IL 60637, USA}

\author{Xuntao Wu}
\affiliation{Pritzker School of Molecular Engineering, University of Chicago, Chicago IL 60637, USA}

\author{Ming-Han Chou}
\altaffiliation[Present address: ]{AWS Center for Quantum Computing, Pasadena, CA 91125, USA}
\affiliation{Department of Physics, University of Chicago, Chicago IL 60637, USA}

\author{Gustav Andersson}
\affiliation{Pritzker School of Molecular Engineering, University of Chicago, Chicago IL 60637, USA}

\author{Christopher R. Conner}
\affiliation{Pritzker School of Molecular Engineering, University of Chicago, Chicago IL 60637, USA}

\author{Yash J. Joshi}
\affiliation{Pritzker School of Molecular Engineering, University of Chicago, Chicago IL 60637, USA}

\author{Shiheng Li}
\affiliation{Department of Physics, University of Chicago, Chicago IL 60637, USA}

\author{Jacob M. Miller}
\affiliation{Department of Physics, University of Chicago, Chicago IL 60637, USA}

\author{Rhys G. Povey}
\altaffiliation[Present address: ]{Institute for Quantum Optics and Quantum Information, Austrian Academy of Sciences, A-1090 Vienna, Austria}
\affiliation{Department of Physics, University of Chicago, Chicago IL 60637, USA}

\author{Hong Qiao}
\affiliation{Pritzker School of Molecular Engineering, University of Chicago, Chicago IL 60637, USA}

\author{Eric Chitambar}
\affiliation{Department of Electrical and Computer Engineering, University of Illinois at Urbana-Champaign, Urbana IL 61801, USA}

\author{Andrew N. Cleland}
\email{anc@uchicago.edu}
\affiliation{Pritzker School of Molecular Engineering, University of Chicago, Chicago IL 60637, USA}
\affiliation{Center for Molecular Engineering and Material Science Division, Argonne National Laboratory, Lemont IL 60439, USA}

\date{\today}

\begin{abstract}
We present a three-node quantum communication testbed with a triangular topology, each side of the triangle formed by a 1.3-meter-long transmission line. We demonstrate state transfer and entanglement generation between any two nodes, generate genuine multipartite entangled GHZ states, and implement quantum secret sharing (QSS) of classical information. Our experiments show that the QSS protocol can effectively detect eavesdropping, ensuring the secure sharing of secrets. This device forms a testbed for robust and secure quantum networking, enabling testing of more advanced quantum communication protocols. 
\end{abstract}

\keywords{Superconducting Qubit, Quantum Network, Quantum Secret Sharing}
\maketitle
\textit{Introduction.}---Quantum networks composed of interconnected quantum processors \cite{Kimble2008, Wehner2018, Pompili2021, Knaut2024, Liu2024} enable a number of useful quantum applications, including quantum key distribution \cite{Bennett1984}, distributed quantum computing \cite{Gottesman1999, Jiang2007, Monroe2014}, and distributed quantum sensing \cite{Proctor2018, Zhang2021}.

Superconducting qubits, meanwhile, provide a promising platform for quantum information processing. Quantum networks of superconducting qubits can enhance the scalability of superconducting quantum processors \cite{Bravyi2022} and further serve as testbeds for quantum communication protocols. Recent experimental advances in quantum communication include quantum state transfer and entanglement generation \cite{Axline2018, Kurpiers2018, CampagneIbarcq2018, Leung2019, Zhong2019, Chang2020, Burkhart2021, Zhong2021, Storz2023, Grebel2024, Niu2023}, entanglement distillation \cite{Yan2022}, and gate teleportation \cite{Chou2018, Qiu2025} between remote superconducting qubits.

In this Letter, we describe a three-node superconducting quantum network with a triangular topology, where each node comprises two superconducting qubits capacitively coupled to one another, and the communication links between each node are 1.3~m in length. We first demonstrate state transfer and entanglement generation between each pair of nodes, achieving an average transfer efficiency of $(88.8\pm 1.2)\%$ and Bell state fidelity of $(95.2\pm 1.1)\%$. Using the generated Bell pairs, we perform entanglement swapping \cite{Zukowski1993}, resulting in swapped Bell states with an average fidelity of $(80.4\pm 1.3)\%$. We then deterministically extend this to genuine multipartite entanglement (GME) across three nodes, by generating a GHZ-3 state with fidelity $(82.6\pm 0.8)\%$, and a GHZ-5 state with fidelity $(70.3\pm 1.1)\%$, both above the GME fidelity threshold of $50\%$ \cite{Guehne2010}. We then use the three-qubit entanglement to demonstrate quantum secret sharing (QSS) \cite{ Hillery1999, Gottesman2000}, achieving an error rate of $(21.5\pm 1.1)\%$, below the local-theory threshold of $25\%$ \cite{Tittel2001}. The QSS protocol thus allows us to detect external eavesdropping by monitoring the key error rate. This three-node quantum network testbed enables testing of more sophisticated quantum communication protocols.

\begin{figure}
    \centering
    \includegraphics{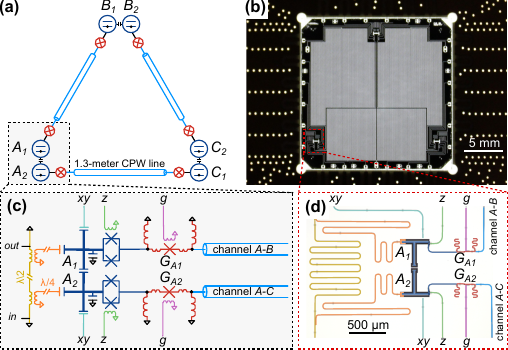}
    \caption{Experimental configuration. (a) Schematic of triangular communication network: Each of the three nodes contains two coupled qubits (dark blue). Each qubit is connected to a 1.3-meter-long CPW transmission line (light blue) via a tunable coupler (red), and to its neighboring qubit through a capacitor (black). The three 1.3-meter-long transmission lines form the edges of the triangle. (b) Backside-illuminated photograph of a fabricated device, wire-bonded to a circuit board for measurement. Nodes are bottom left, bottom right, and top center; most of the device is occupied by the 1.3-meter-long meandered coplanar waveguide transmission lines. (c) Detailed circuit and (d) physical layout of node $A$: Two capacitively coupled Xmon qubits $A_{1,2}$ (dark blue) are each individually coupled to an $xy$ drive line (cyan) and a $z$ flux line (green), and to a communication channel (light blue, $A-B$ and $A-C$) through a tunable coupler $G_{A1, A2}$ (red), with the coupling strength tuned by a $g$ flux line (purple). Orange and sepia traces are quarter-wave readout resonators and Purcell filters, respectively.}
    \label{fig1}
\end{figure}

\textit{Device.}---A schematic of the device is shown in Fig.~\ref{fig1}(a), with a backside-illuminated image of the device in Fig.~\ref{fig1}(b). The three nodes, labeled $A$, $B$, and $C$, are connected in a triangular geometry. Each node comprises two capacitively-coupled Xmon qubits $Q_i$ \cite{Koch2007, Barends2013}, with $Q=A,B,C$ and $i=1,2$. Each qubit $Q_i$ is connected through a tunable coupler $G_{Qi}$ \cite{Chen2014} to a transmission line. These lines are 1.3-meter-long coplanar waveguides (CPWs; meandered structures in Fig.~\ref{fig1}(b)), with a characteristic impedance of $50~\mathrm{\Omega}$.

The circuit and layout for node $A$ are displayed in Figs.~\ref{fig1}(c) and (d), respectively. Each Xmon qubit $A_i$ (dark blue) is individually controlled by an $xy$ drive line for $X$ and $Y$ rotations (cyan), and a $z$ flux line for frequency tuning (green), which yields effective $Z$ rotations in the system frame. Readout is performed via a dispersively-coupled $\lambda/4$ resonator (orange). Each node includes a $\lambda/2$ single-stage bandpass Purcell filter (sepia) to protect the qubits from decay and noise via the readout channel \cite{Jeffrey2014, Yan2023}. The couplers $G_{Ai}$ (red) are controlled by $g$ flux lines (purple) to adjust the coupling strength. The 1.3-meter-long CPW transmission lines (light blue) are galvanically connected to the couplers. 

Devices are fabricated on 100~mm-diameter sapphire wafers, including many duplicate and closely similar layouts. The wafer is then singulated into $20~\mathrm{mm} \times 20~\mathrm{mm}$ dies. The fabrication process is described in more detail in the Supplemental Material~\cite{supp}. A single die is wire-bonded to a connectorized gold-plated printed circuit board, placed in a metal magnetically-shielded enclosure, and cooled to $10~\mathrm{mK}$ on the mixing chamber of a dilution refrigerator for measurement. The wiring and measurement setup are provided in the Supplemental Material \cite{supp}. 

We calibrate the device by first turning off the adjustable couplers and characterizing the qubits individually. Each qubit exhibits an energy relaxation time $T_1$ of approximately $10~\mathrm{\mu s}$ and a pure dephasing time $T_\phi$ of around $2~\mathrm{\mu s}$. Single-qubit rotations are performed using $30~\mathrm{ns}$ on-resonance microwave pulses with a cosine envelope and DRAG correction \cite{Motzoi2009}. We benchmark all the single-qubit gates with randomized benchmarking \cite{Knill2008} and measure an average gate fidelity of around $99.6\%$, limited by qubit coherence times. 

The coupled qubits in each node can perform two-qubit operations, mediated by their coupling capacitor, including $\mathrm{CZ}$ and $\mathrm{iSWAP}$ gates, by tuning the qubit frequencies. We characterize the $\mathrm{CZ}$ gates using cross-entropy benchmarking \cite{Boixo2018}, measuring an average gate fidelity of approximately $95.6\%$, limited primarily by qubit dephasing due to the relatively short $T_2$ times. Detailed qubit parameters and benchmarking results are provided in the Supplemental Material \cite{supp}.

\begin{figure}[!tb]
    \centering
    \includegraphics{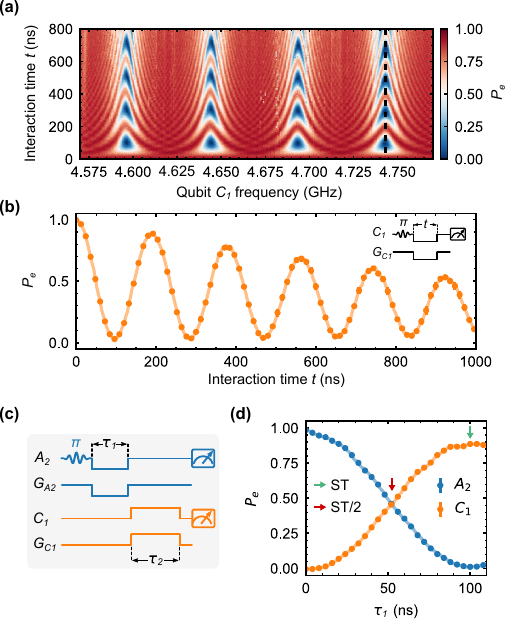}
    \caption{Two-node state transfer and entanglement generation. (a) Vacuum Rabi oscillations between qubit $C_1$ and multiple standing modes in the transmission line linking nodes $A$ and $C$, displaying the $50~\mathrm{MHz}$ free-spectral range of the 1.3-meter-long CPW line. (b) Evolution of populations in $C_1$ when interacting with the $4.743~\mathrm{GHz}$ CPW mode (dashed line in (a)), with the pulse sequence shown in the inset. (c) Pulse sequence for state transfer and entanglement generation between nodes $A$ and $C$. First, qubit $A_2$ is placed in its excited state $|e\rangle$ with a tuned $\pi$ pulse. The coupler $G_{A2}$ is then turned on and $A_2$ is brought on resonance with the $4.743~\mathrm{GHz}$ CPW mode for a duration $\tau_1$. The excitation is subsequently swapped from the CPW mode to $C_1$ by turning on the coupler $G_{C1}$ and bringing $C_1$ on resonance with the mode for a duration $\tau_2=\pi/2g_2$. (d) The $|e\rangle$ state populations in $A_2$ (blue) and $C_1$ (orange) versus first swap duration $\tau_1$. The green and red arrows indicate the times $\tau_1$ corresponding to state transfer (ST) and Bell state generation (ST/2), respectively.}\label{fig2}
\end{figure}

We characterize the communication channels using the qubits, as shown in Fig.~\ref{fig2}. The pulse sequence is shown in the inset in Fig.~\ref{fig2}(b): We excite the emitting qubit to its excited state $|e\rangle$ with a $\pi$ pulse, then turn on the coupler to the CPW transmission line and simultaneously tune the qubit frequency to the desired CPW mode.  Fig.~\ref{fig2}(a) displays the resulting vacuum Rabi oscillations between qubit $C_1$ as a function of time (vertical) and frequency (horizontal), the chevron patterns in the excitation probability $P_e$ of $C_1$ each indicating a separate communication mode, revealing the channel's expected free spectral range (FSR) $\omega_{\mathrm{FSR}}/2\pi$ of $50~\mathrm{MHz}$. In Fig.~\ref{fig2}(b), we show the evolution of the qubit excitation probability when interacting with the $4.743~\mathrm{GHz}$ standing mode (dashed line in Fig.~\ref{fig2}(a)), which we designate as the communication mode $M$. The lifetime $T_1$ and dephasing time $T_2$ of mode $M$ are measured by swapping the qubit excitation into the mode, and monitoring the communication mode decay. We measure a mode $M$ energy lifetime $T_1 = 1.2~\mathrm{\mu s}$ and phase coherence time $T_2 = 2.3~\mathrm{\mu s}$, corresponding to a mode quality factor of $3\times 10^4$. Detailed results for each of the three communication channels are provided in the Supplemental Material \cite{supp}. The maximum coupling strength between the qubits and each communication mode $g_\mathrm{max}/2\pi$ is $10~\mathrm{MHz}$.

\textit{State transfer and entanglement generation.}---We demonstrate state transfer and entanglement generation between nodes $A$ and $C$ using the pulse sequence shown in Fig.~\ref{fig2}(c). First, qubit $A_2$ is placed in its $|e\rangle$ state, then tuned on-resonance with the communication mode $M$ with its coupler $G_{A2}$ turned on, for a duration $\tau_1$. This results in the state $|A_2 M \rangle = \cos(g_1\tau_1) |e0\rangle + i \sin(g_1\tau_1) |g1\rangle$, where $g_1$ is the coupling strength between $A_2$ and $M$. Next, we turn on the coupler $G_{C1}$ on the opposite end of the CPW line and bring qubit $C_1$ on resonance with the communication mode $M$ for a duration of $\tau_2=\pi/2g_2$, where $g_2$ is the coupling strength between $C_1$ and the communication mode $M$. This swaps states between $M$ and $C_1$, resulting in the state $|A_2C_1\rangle = \cos(g_1\tau_1) |eg\rangle - \sin(g_1\tau_1) |ge\rangle$. To achieve the best Bell state fidelity and state transfer efficiency, we set the couplings $g_1/2\pi$ to $2.5~\mathrm{MHz}$ to minimize the leakage to the other adjacent standing modes in the transmission line, and $g_2/2\pi$ to $3.1~\mathrm{MHz}$ to minimize the second swap time.

In Fig.~\ref{fig2}(d), we plot the population of the qubit in the $|e\rangle$ state versus the first swap duration $\tau_1$. When $\tau_1=\pi/4g_1=50~\mathrm{ns}$ (indicated by the red arrow), a Bell state $|\psi^-\rangle=(|eg\rangle-|ge\rangle)/\sqrt{2}$ is generated, a process labeled as ST/2. The fidelity $\mathcal{F}_{\mathrm{Bell}}=\langle\psi^-|\rho_{A_2C_1}|\psi^-\rangle$ of the generated Bell state $\rho_{A_2C_1}$ between $A_2$ and $C_1$ is $95.1\pm0.8\%$, consistent with numerical simulations, which give a fidelity of $94.3\%$ \cite{supp}. The state fidelity is primarily limited by qubit dephasing. When $\tau_1=\pi/2g_1=100~\mathrm{ns}$ (indicated by the green arrow), the state is fully transferred from $A_2$ to $C_1$, a process we label as ST. The state transfer efficiency $\eta_t$ between $A_2$ and $C_1$ is $(88.7\pm 0.8)\%$, defined as the transferred $|e\rangle$ state population. This fidelity is limited primarily by channel loss.

The benchmarked state transfer efficiency and Bell state fidelity for all three channels are listed in the Supplemental Material \cite{supp}. By generating two Bell pairs and using local qubit operations, we demonstrate entanglement swapping \cite{Zukowski1993}, achieving an average swapped Bell state fidelity of $80.4\pm 1.3\%$, limited primarily by qubit dephasing and readout errors. Further details are provided in the Supplemental Material \cite{supp}.

\begin{figure}
    \centering
    \includegraphics{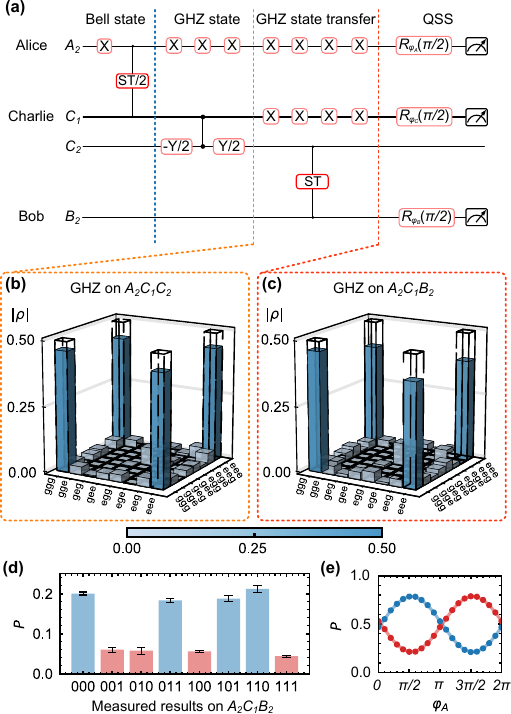}
    \caption{Quantum secret sharing. (a) Pulse sequence for quantum secret sharing. We first generate a Bell state between $A_2$ and $C_1$ using the procedure shown in Fig.~\ref{fig2}(c), which we then extend to a GHZ-3 state between $A_2C_1C_2$, by applying a CNOT gate between $C_1$ and $C_2$. Next we transfer $C_2$ to $B_2$, resulting in a GHZ state in $A_2C_1B_2$ (Alice, Charlie and Bob, respectively). Repetitive dynamical decoupling $X$ gates are applied to the idling qubits during this process. Finally, single-qubit rotations are applied to enable measuring the qubit states on the $x$ or $y$ axes of the Bloch sphere. (b) Density matrices for the $A_2C_1C_2$ and (c) $A_2C_1B_2$ GHZ states. Dashed lines represent ideal GHZ states. (d) Measured probabilities for different three-qubit states when the three qubits are measured along the $x$ axis. Blue bars indicate results where Bob and Charlie can successfully decode the secret held by Alice. (e) Probabilities of measuring $A_2C_1B_2$ in the states $|000\rangle$, $|011\rangle$, $|101\rangle$, or $|110\rangle$ (blue, as in (d)), or in $|001\rangle$, $|010\rangle$, $|100\rangle$, or $|111\rangle$ (red) versus the rotation $\varphi_A$ applied to qubit $A_2$.}\label{fig3}
\end{figure}

\textit{Quantum secret sharing using {GHZ} states.}--- A typical and natural application of a multipartite communication network is secret sharing. Classical secret sharing refers to methods for distributing a secret, such as a bit string, among a group of people \cite{Shamir1979, Blakley1979}. In a $(k, n)$ threshold scheme, a secret shared among $n$ people can only be reconstructed when a group of $k$ or more people combine their pieces of information. If fewer than $k$ people coordinate, no information about the secret is leaked to the cohort. Quantum secret sharing provides additional security by using quantum entanglement and the no-cloning theorem. Various QSS protocols exist \cite{Cleve1999, Hillery1999, Gottesman2000, Karlsson1999, Guo2003, Xiao2004, Zhang2005, Zhang2005a, Markham2008}, and QSS has been demonstrated experimentally with optics \cite{Tittel2001,Chen2005,Gaertner2007,Bogdanski2008,Bell2014,Xiao2024,Zhang2024,Wang2024,Zhang2025} and using locally-connected superconducting qubits \cite{Basak2023,graves2024}. Here we demonstrate QSS of classical information using our triangular quantum network.

The simplest case of QSS involves Alice sharing a secret (a bit) with Bob and Charlie using a GHZ state $|\psi\rangle_{\mathrm{GHZ}}=(|ggg\rangle +|eee\rangle)/\sqrt{2}$ that is shared among the three of them \cite{Hillery1999, Greenberger1989}. Each participant then measures their qubit along the $x$ or $y$ axis randomly, and communicates their chosen axis (but not their measurement result) through a classical, public channel. The outcome of Alice's measurement is the secret bit that is known to her and also shared between Bob and Charlie. By combining their measurement results, Bob and Charlie can deduce what Alice measured \cite{Hillery1999}. However, neither Bob nor Charlie by themselves can extract any information about the secret Alice holds.

We use the pulse sequence in Fig.~\ref{fig3}(a) to generate a GHZ state across the three nodes. First, we generate a Bell state $|\psi\rangle=(|ge\rangle-|eg\rangle)/\sqrt{2}$ between $A_2$ and $C_1$ using the ST/2 process described in Figs.~\ref{fig2}(c) and (d). By applying a CNOT gate (composed of two single-qubit gates on $C_2$ and a CZ gate between $C_1$ and $C_2$) and three $X$ gates on $A_2$, we obtain a GHZ state $|A_2C_1C_2\rangle=(|ggg\rangle-|eee\rangle)/\sqrt{2}$ with a fidelity of $(88.6\pm 0.7)\%$. The density matrix is plotted in Fig.~\ref{fig3}(b). The three $X$ gates act as a dynamical decoupling sequence to reduce the loss of fidelity from the dephasing of $A_2$.

Next, we transfer the state of $C_2$ to $B_1$ using the ST process described in Figs.~\ref{fig2}(c) and (d), resulting in a GHZ state shared by the three nodes, $|A_2C_1B_1\rangle=(|ggg\rangle+|eee\rangle)/\sqrt{2}$. The phase flip appears because the state transfer process is equivalent to two $\mathrm{iSWAP}$ gates. We also apply four dynamical decoupling $X$ gates to $A_2$ and $C_1$ during the transfer process to mitigate the dephasing effect. We achieve a GHZ state fidelity of $(82.6\pm 0.8)\%$ with the density matrix plotted in Fig.~\ref{fig3}(c). We observe an improvement of approximately $5\%$ in state fidelity compared to omitting the $X$ decoupling gates. 

We can further expand this GHZ state to five qubits, achieving a state fidelity of $(70.3\pm 1.1)\%$, which is above the fidelity threshold of $50\%$ for genuine multipartite entanglement \cite{Guehne2010}. More details on the GHZ state generation can be found in the Supplemental Material \cite{supp}. 

To perform the QSS measurements along the $x$ or $y$ axes, we apply single-qubit $\pi/2$ rotations $R_\varphi(\pi/2)$ before measuring each qubit along its $z$ axis, where $\varphi$ is the angle of the rotation axis with respect to the $x$-axis on the $xy$ plane of the Bloch sphere. When $\varphi=0$ ($\varphi=\pi/2$), the post-rotation measurement is along the $y$ ($x$) axis. The state $|g\rangle$ gives a measurement result of 0 and $|e\rangle$ gives 1. In Fig.~\ref{fig3}(d), we plot the probability distribution of measurement results for $A_2C_1B_2$ measured along the $x$ axis ($\varphi_A=\varphi_B=\varphi_C=\pi/2$). When the measurement results of Bob ($B_2$) and Charlie ($C_1$) are the same (different), the bit held by Alice ($A_2$) should be 0 (1), as shown by the blue bars. Ideally, these measured probabilities should all be $25\%$, while those in red should be zero. We observe an average probability of $(19.6\pm 1.1)\%$ for a correct encoding of Alice's measurement outcome, leading to an overall QSS error rate of $(21.5\pm 1.1)\%$. This is below the $25\%$ threshold corresponding to a naive intercept-and-measure attack by a cheating Bob or Charlie \cite{Hillery1999}, and which also corresponds to the noise threshold for which the GHZ state no longer violates the Mermin-Bell inequality \cite{Tittel2001}. The errors are primarily due to imperfect GHZ state generation and imperfect readout. 

QSS ensures that Bob or Charlie cannot extract the secret individually, a restriction that we experimentally verify. Here we take Bob as an example and assume Alice measures along $x$ axis. Suppose now that Bob deviates from the protocol and tries to learn the value of Alice's measurement outcome (i.e. the secret) after knowing her measurement direction by measuring his system. 
His optimal probability of correctly guessing the secret, $p_B$, is given by \cite{Barnett2009} $p_B = \frac{1}{2}+\frac{1}{2}\Vert p_{A=0}\rho_{B|A=0}-p_{A=1}\rho_{B|A=1}\Vert$, where $\Vert \cdot\Vert$ is the trace norm, $p_{A=i}$ is the probability of Alice measuring $i$ along $x$ axis and $\rho_{B|A=i}$ is the condition matrix of Bob when Alice measures $i$. From the tomography data, we obtain $p_B=54.2\%$, which is close to the ideal probability of $50\%$. 

By sweeping $\varphi_A$, we observe that the sum of probabilities for measuring $000$, $011$, $101$, and $110$ in $A_2C_1B_2$ changes in proportion to $1+\cos\varphi_A$, as shown in Fig.~\ref{fig3}(e). When $\varphi_A=0$, corresponding to measurement along Alice's $y$ axis, the probability sum becomes $50\%$, indicating that Bob and Charlie cannot extract any information about the secret. Bob and Charlie can deduce what Alice measured half of the time because everyone selects the axes randomly at the beginning \cite{Hillery1999}. This emphasizes that each participant needs to communicate the measurement axes they used through a classical channel, to discard such useless (e.g. when $\varphi_A=0$ and $\varphi_B=\phi_C=\pi/2$) results.

\begin{figure}[tb]
    \centering
    \includegraphics{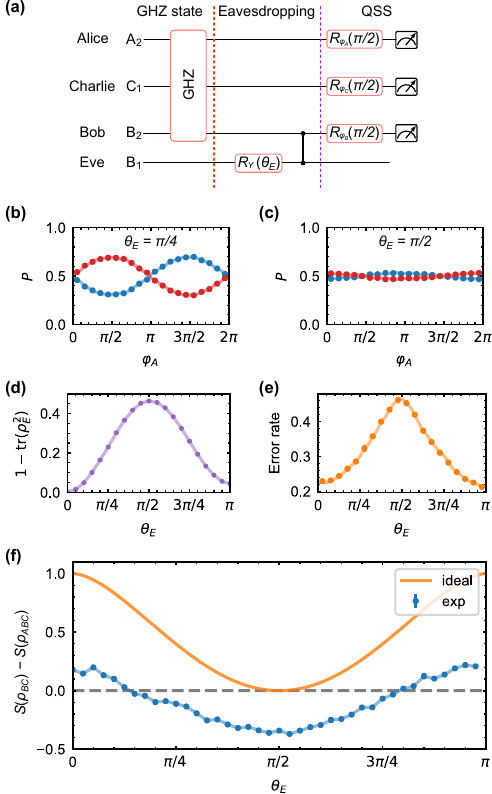}
    \caption{Detection of eavesdropping. (a) Pulse sequence for the eavesdropper Eve ($B_1$, adjacent to Bob $B_2$) to perform an entangle-and-measure attack. Eve performs a single-qubit rotation $R_Y(\theta_E)$, followed by a CZ gate between Bob ($B_2$) and Eve ($B_1$). (b) Probabilities of measuring $A_2C_1B_2$ in 000, 011, 101, 110 (blue) and 001, 010, 100, 111 (red) versus $\varphi_A$ for $\theta_E=\pi/4$ and (c) $\theta_E=\pi/2$. (d) Linear entropy of Eve ($B_1$) versus $\theta_E$. (e) QSS error rate versus $\theta_E$. (f) The lower limit of $P$ versus $\theta_E$. Blue dots represent experimental data, while the orange line indicates the ideal case.}\label{fig4}
\end{figure}

QSS also offers the ability to detect eavesdroppers. If an eavesdropper, Eve, attempts to steal information from Alice, Bob, and Charlie, she cannot directly copy the state from any party due to the no-cloning theorem \cite{Park1970}. However, she might use strategies such as intercept-and-resend or entangle-and-measure to gain partial information \cite{Basak2023}.

Here we implement an entangle-and-measure attack, where Eve tries to steal the secret by entangling herself with Bob. We demonstrate that the QSS protocol can detect such an attack, ensuring that Eve cannot gain more information about the secret than Bob and Charlie, and further she reveals her presence.

We use the pulse sequence shown in Fig.~\ref{fig4}(a) for Eve to perform the entangle-and-measure attack. First, we perform a rotation $R_Y(\theta_E)$ along Eve's $y$ axis, which results in Eve's state $|\psi\rangle_E = \cos(\theta_E/2) |g\rangle - \sin(\theta_E/2) |e\rangle$. Then we perform a $\mathrm{CZ}$ gate as an entangling operation between Eve and Bob. Following this, if $\theta_E=0$ or $\pi$, Eve is in $|g\rangle$ or $|e\rangle$ and not entangled with Alice, Bob, or Charlie. When instead $0<\theta_E<\pi$, the four qubits are entangled. Notably, when $\theta_E = \pi/2$, the entangling operation functions as part of a $\mathrm{CNOT}$ gate. This entanglement causes the reduced state of $A_2C_1B_2$ to become mixed even in the ideal case, reducing its fidelity with respect to the ideal GHZ state by a factor $(1+|\cos \theta_E|)/2$ when tracing out Eve's state. We plot the density matrices of $A_2C_1B_2$ for $\theta_E=\pi/4$ and $\pi/2$ in the Supplemental Material \cite{supp}. 

To illustrate how entangled Eve is with the other three parties, we plot the linear entropy (impurity) of Eve's state $1-\mathrm{tr}(\rho_E^2)$, in Fig.~\ref{fig4}(d), where $\rho_E$ is Eve's reduced density matrix. After Eve finishes the entangling operation, Alice, Bob, and Charlie perform the QSS protocol as described above, where the QSS error rate becomes $(1-|\cos\theta_E|)/4$. In Figs.~\ref{fig4}(b) and (c), we plot the same quantities as in Fig.~\ref{fig3}(e) for $\theta_E=\pi/4$ and $\pi/2$. We observe a decrease in visibility as $\theta_E$ increases from $0$ to $\pi/2$. In Fig.~\ref{fig4}(e), we plot the extracted QSS error rate versus $\theta_E$. This shows that as Eve's entanglement increases, allowing her to acquire more information, the error rate also increases. Consequently, by monitoring the error rate, Alice, Bob, and Charlie can detect the presence of an eavesdropper.

Despite Eve gaining some information about Alice's secret, we observe that it is sometimes possible to asymptotically suppress this information, even with the experimental noise in our setup. The celebrated Devetak-Winter theorem ensures that a private key can be obtained asymptotically between $A$ and $BC$ (working jointly) whenever $P=I(X:BC)-I(X:E)> 0$ \cite{Devetak2004}. Here, $E$ is Eve's system, while $X$ is the classical variable generated when Alice measures her system in some basis, and $I$ denotes the mutual information. One can show that $P\geq S(\rho_{BC})-S(\rho_{ABC})$ where $S$ is the von Neumann entropy (see the Supplemental Material \cite{supp}). In an ideal implementation of this specific entangle-and-measure attack, the coherent information $P$ is always positive (orange line). In practice, the lower limit becomes negative for some values of $\theta_E$ because other loss channels cause additional errors.  Nevertheless, for some non-optimal attacks, the key rate is still positive.  Notably, when $\theta_E=0$, the key can still be generated under the uncharacterized and untrusted noise of our setup.



\textit{Discussion and outlook.}--- We have designed and built a three-node superconducting quantum communication testbed with a triangular topology. We use this setup to demonstrate state transfer and entanglement generation between any two nodes, enabling the distribution of genuine multipartite entangled GHZ states and the implementation of quantum secret sharing. Our results reinforce that quantum secret sharing can detect eavesdropping, ensuring the security of communication. This testbed paves the way for demonstrating more advanced quantum communication protocols, such as multipartite entanglement purification \cite{Duer2003, Murao1998}, quantum voting \cite{Hillery2006, Xue2017}, the quantum Byzantine agreement \cite{Lamport1982}, and can also be used to test network nonlocality \cite{Renou2019, Tavakoli2022}.

\bibliography{QSS_ref}

\begin{thebibliography}{70}%
\makeatletter
\providecommand \@ifxundefined [1]{%
 \@ifx{#1\undefined}
}%
\providecommand \@ifnum [1]{%
 \ifnum #1\expandafter \@firstoftwo
 \else \expandafter \@secondoftwo
 \fi
}%
\providecommand \@ifx [1]{%
 \ifx #1\expandafter \@firstoftwo
 \else \expandafter \@secondoftwo
 \fi
}%
\providecommand \natexlab [1]{#1}%
\providecommand \enquote  [1]{``#1''}%
\providecommand \bibnamefont  [1]{#1}%
\providecommand \bibfnamefont [1]{#1}%
\providecommand \citenamefont [1]{#1}%
\providecommand \href@noop [0]{\@secondoftwo}%
\providecommand \href [0]{\begingroup \@sanitize@url \@href}%
\providecommand \@href[1]{\@@startlink{#1}\@@href}%
\providecommand \@@href[1]{\endgroup#1\@@endlink}%
\providecommand \@sanitize@url [0]{\catcode `\\12\catcode `\$12\catcode
  `\&12\catcode `\#12\catcode `\^12\catcode `\_12\catcode `\%12\relax}%
\providecommand \@@startlink[1]{}%
\providecommand \@@endlink[0]{}%
\providecommand \url  [0]{\begingroup\@sanitize@url \@url }%
\providecommand \@url [1]{\endgroup\@href {#1}{\urlprefix }}%
\providecommand \urlprefix  [0]{URL }%
\providecommand \Eprint [0]{\href }%
\providecommand \doibase [0]{https://doi.org/}%
\providecommand \selectlanguage [0]{\@gobble}%
\providecommand \bibinfo  [0]{\@secondoftwo}%
\providecommand \bibfield  [0]{\@secondoftwo}%
\providecommand \translation [1]{[#1]}%
\providecommand \BibitemOpen [0]{}%
\providecommand \bibitemStop [0]{}%
\providecommand \bibitemNoStop [0]{.\EOS\space}%
\providecommand \EOS [0]{\spacefactor3000\relax}%
\providecommand \BibitemShut  [1]{\csname bibitem#1\endcsname}%
\let\auto@bib@innerbib\@empty
\bibitem [{\citenamefont {Kimble}(2008)}]{Kimble2008}%
  \BibitemOpen
  \bibfield  {author} {\bibinfo {author} {\bibfnamefont {H.~J.}\ \bibnamefont
  {Kimble}},\ }\bibfield  {title} {\bibinfo {title} {The quantum internet},\
  }\href {https://doi.org/10.1038/nature07127} {\bibfield  {journal} {\bibinfo
  {journal} {Nature}\ }\textbf {\bibinfo {volume} {453}},\ \bibinfo {pages}
  {1023} (\bibinfo {year} {2008})}\BibitemShut {NoStop}%
\bibitem [{\citenamefont {Wehner}\ \emph {et~al.}(2018)\citenamefont {Wehner},
  \citenamefont {Elkouss},\ and\ \citenamefont {Hanson}}]{Wehner2018}%
  \BibitemOpen
  \bibfield  {author} {\bibinfo {author} {\bibfnamefont {S.}~\bibnamefont
  {Wehner}}, \bibinfo {author} {\bibfnamefont {D.}~\bibnamefont {Elkouss}},\
  and\ \bibinfo {author} {\bibfnamefont {R.}~\bibnamefont {Hanson}},\
  }\bibfield  {title} {\bibinfo {title} {Quantum internet: {A} vision for the
  road ahead},\ }\href {https://doi.org/10.1126/science.aam9288} {\bibfield
  {journal} {\bibinfo  {journal} {Science}\ }\textbf {\bibinfo {volume}
  {362}},\ \bibinfo {pages} {eaam9288} (\bibinfo {year} {2018})}\BibitemShut
  {NoStop}%
\bibitem [{\citenamefont {Pompili}\ \emph {et~al.}(2021)\citenamefont
  {Pompili}, \citenamefont {Hermans}, \citenamefont {Baier}, \citenamefont
  {Beukers}, \citenamefont {Humphreys}, \citenamefont {Schouten}, \citenamefont
  {Vermeulen}, \citenamefont {Tiggelman}, \citenamefont {dos Santos~Martins},
  \citenamefont {Dirkse}, \citenamefont {Wehner},\ and\ \citenamefont
  {Hanson}}]{Pompili2021}%
  \BibitemOpen
  \bibfield  {author} {\bibinfo {author} {\bibfnamefont {M.}~\bibnamefont
  {Pompili}}, \bibinfo {author} {\bibfnamefont {S.~L.~N.}\ \bibnamefont
  {Hermans}}, \bibinfo {author} {\bibfnamefont {S.}~\bibnamefont {Baier}},
  \bibinfo {author} {\bibfnamefont {H.~K.~C.}\ \bibnamefont {Beukers}},
  \bibinfo {author} {\bibfnamefont {P.~C.}\ \bibnamefont {Humphreys}}, \bibinfo
  {author} {\bibfnamefont {R.~N.}\ \bibnamefont {Schouten}}, \bibinfo {author}
  {\bibfnamefont {R.~F.~L.}\ \bibnamefont {Vermeulen}}, \bibinfo {author}
  {\bibfnamefont {M.~J.}\ \bibnamefont {Tiggelman}}, \bibinfo {author}
  {\bibfnamefont {L.}~\bibnamefont {dos Santos~Martins}}, \bibinfo {author}
  {\bibfnamefont {B.}~\bibnamefont {Dirkse}}, \bibinfo {author} {\bibfnamefont
  {S.}~\bibnamefont {Wehner}},\ and\ \bibinfo {author} {\bibfnamefont
  {R.}~\bibnamefont {Hanson}},\ }\bibfield  {title} {\bibinfo {title}
  {Realization of a multinode quantum network of remote solid-state qubits},\
  }\href {https://doi.org/10.1126/science.abg1919} {\bibfield  {journal}
  {\bibinfo  {journal} {Science}\ }\textbf {\bibinfo {volume} {372}},\ \bibinfo
  {pages} {259} (\bibinfo {year} {2021})}\BibitemShut {NoStop}%
\bibitem [{\citenamefont {Knaut}\ \emph {et~al.}(2024)\citenamefont {Knaut},
  \citenamefont {Suleymanzade}, \citenamefont {Wei}, \citenamefont {Assumpcao},
  \citenamefont {Stas}, \citenamefont {Huan}, \citenamefont {Machielse},
  \citenamefont {Knall}, \citenamefont {Sutula}, \citenamefont {Baranes},
  \citenamefont {Sinclair}, \citenamefont {De-Eknamkul}, \citenamefont
  {Levonian}, \citenamefont {Bhaskar}, \citenamefont {Park}, \citenamefont
  {Lončar},\ and\ \citenamefont {Lukin}}]{Knaut2024}%
  \BibitemOpen
  \bibfield  {author} {\bibinfo {author} {\bibfnamefont {C.~M.}\ \bibnamefont
  {Knaut}}, \bibinfo {author} {\bibfnamefont {A.}~\bibnamefont {Suleymanzade}},
  \bibinfo {author} {\bibfnamefont {Y.-C.}\ \bibnamefont {Wei}}, \bibinfo
  {author} {\bibfnamefont {D.~R.}\ \bibnamefont {Assumpcao}}, \bibinfo {author}
  {\bibfnamefont {P.-J.}\ \bibnamefont {Stas}}, \bibinfo {author}
  {\bibfnamefont {Y.~Q.}\ \bibnamefont {Huan}}, \bibinfo {author}
  {\bibfnamefont {B.}~\bibnamefont {Machielse}}, \bibinfo {author}
  {\bibfnamefont {E.~N.}\ \bibnamefont {Knall}}, \bibinfo {author}
  {\bibfnamefont {M.}~\bibnamefont {Sutula}}, \bibinfo {author} {\bibfnamefont
  {G.}~\bibnamefont {Baranes}}, \bibinfo {author} {\bibfnamefont
  {N.}~\bibnamefont {Sinclair}}, \bibinfo {author} {\bibfnamefont
  {C.}~\bibnamefont {De-Eknamkul}}, \bibinfo {author} {\bibfnamefont {D.~S.}\
  \bibnamefont {Levonian}}, \bibinfo {author} {\bibfnamefont {M.~K.}\
  \bibnamefont {Bhaskar}}, \bibinfo {author} {\bibfnamefont {H.}~\bibnamefont
  {Park}}, \bibinfo {author} {\bibfnamefont {M.}~\bibnamefont {Lončar}},\ and\
  \bibinfo {author} {\bibfnamefont {M.~D.}\ \bibnamefont {Lukin}},\ }\bibfield
  {title} {\bibinfo {title} {Entanglement of nanophotonic quantum memory nodes
  in a telecom network},\ }\href {https://doi.org/10.1038/s41586-024-07252-z}
  {\bibfield  {journal} {\bibinfo  {journal} {Nature}\ }\textbf {\bibinfo
  {volume} {629}},\ \bibinfo {pages} {573} (\bibinfo {year}
  {2024})}\BibitemShut {NoStop}%
\bibitem [{\citenamefont {Liu}\ \emph {et~al.}(2024)\citenamefont {Liu},
  \citenamefont {Luo}, \citenamefont {Yu}, \citenamefont {Wang}, \citenamefont
  {Wang}, \citenamefont {Hu}, \citenamefont {Li}, \citenamefont {Zheng},
  \citenamefont {Yao}, \citenamefont {Yan}, \citenamefont {Teng}, \citenamefont
  {Jiang}, \citenamefont {Liu}, \citenamefont {Xie}, \citenamefont {Zhang},
  \citenamefont {Mao}, \citenamefont {Jiang}, \citenamefont {Zhang},
  \citenamefont {Bao},\ and\ \citenamefont {Pan}}]{Liu2024}%
  \BibitemOpen
  \bibfield  {author} {\bibinfo {author} {\bibfnamefont {J.-L.}\ \bibnamefont
  {Liu}}, \bibinfo {author} {\bibfnamefont {X.-Y.}\ \bibnamefont {Luo}},
  \bibinfo {author} {\bibfnamefont {Y.}~\bibnamefont {Yu}}, \bibinfo {author}
  {\bibfnamefont {C.-Y.}\ \bibnamefont {Wang}}, \bibinfo {author}
  {\bibfnamefont {B.}~\bibnamefont {Wang}}, \bibinfo {author} {\bibfnamefont
  {Y.}~\bibnamefont {Hu}}, \bibinfo {author} {\bibfnamefont {J.}~\bibnamefont
  {Li}}, \bibinfo {author} {\bibfnamefont {M.-Y.}\ \bibnamefont {Zheng}},
  \bibinfo {author} {\bibfnamefont {B.}~\bibnamefont {Yao}}, \bibinfo {author}
  {\bibfnamefont {Z.}~\bibnamefont {Yan}}, \bibinfo {author} {\bibfnamefont
  {D.}~\bibnamefont {Teng}}, \bibinfo {author} {\bibfnamefont {J.-W.}\
  \bibnamefont {Jiang}}, \bibinfo {author} {\bibfnamefont {X.-B.}\ \bibnamefont
  {Liu}}, \bibinfo {author} {\bibfnamefont {X.-P.}\ \bibnamefont {Xie}},
  \bibinfo {author} {\bibfnamefont {J.}~\bibnamefont {Zhang}}, \bibinfo
  {author} {\bibfnamefont {Q.-H.}\ \bibnamefont {Mao}}, \bibinfo {author}
  {\bibfnamefont {X.}~\bibnamefont {Jiang}}, \bibinfo {author} {\bibfnamefont
  {Q.}~\bibnamefont {Zhang}}, \bibinfo {author} {\bibfnamefont {X.-H.}\
  \bibnamefont {Bao}},\ and\ \bibinfo {author} {\bibfnamefont {J.-W.}\
  \bibnamefont {Pan}},\ }\bibfield  {title} {\bibinfo {title} {Creation of
  memory-memory entanglement in a metropolitan quantum network},\ }\href
  {https://doi.org/10.1038/s41586-024-07308-0} {\bibfield  {journal} {\bibinfo
  {journal} {Nature}\ }\textbf {\bibinfo {volume} {629}},\ \bibinfo {pages}
  {579} (\bibinfo {year} {2024})}\BibitemShut {NoStop}%
\bibitem [{\citenamefont {Bennett}\ and\ \citenamefont
  {Brassard}(1984)}]{Bennett1984}%
  \BibitemOpen
  \bibfield  {author} {\bibinfo {author} {\bibfnamefont {C.~H.}\ \bibnamefont
  {Bennett}}\ and\ \bibinfo {author} {\bibfnamefont {G.}~\bibnamefont
  {Brassard}},\ }\bibfield  {title} {\bibinfo {title} {Quantum cryptography:
  {P}ublic key distribution and coin tossing},\ }in\ \href
  {https://doi.org/10.1016/j.tcs.2014.05.025} {\emph {\bibinfo {booktitle}
  {Proceedings of the IEEE International Conference on Computers, Systems and
  Signal Processing}}}\ (\bibinfo {address} {Bangalore},\ \bibinfo {year}
  {1984})\ pp.\ \bibinfo {pages} {175--179}\BibitemShut {NoStop}%
\bibitem [{\citenamefont {Gottesman}\ and\ \citenamefont
  {Chuang}(1999)}]{Gottesman1999}%
  \BibitemOpen
  \bibfield  {author} {\bibinfo {author} {\bibfnamefont {D.}~\bibnamefont
  {Gottesman}}\ and\ \bibinfo {author} {\bibfnamefont {I.~L.}\ \bibnamefont
  {Chuang}},\ }\bibfield  {title} {\bibinfo {title} {Demonstrating the
  viability of universal quantum computation using teleportation and
  single-qubit operations},\ }\href {https://doi.org/10.1038/46503} {\bibfield
  {journal} {\bibinfo  {journal} {Nature}\ }\textbf {\bibinfo {volume} {402}},\
  \bibinfo {pages} {390} (\bibinfo {year} {1999})}\BibitemShut {NoStop}%
\bibitem [{\citenamefont {Jiang}\ \emph {et~al.}(2007)\citenamefont {Jiang},
  \citenamefont {Taylor}, \citenamefont {Sørensen},\ and\ \citenamefont
  {Lukin}}]{Jiang2007}%
  \BibitemOpen
  \bibfield  {author} {\bibinfo {author} {\bibfnamefont {L.}~\bibnamefont
  {Jiang}}, \bibinfo {author} {\bibfnamefont {J.~M.}\ \bibnamefont {Taylor}},
  \bibinfo {author} {\bibfnamefont {A.~S.}\ \bibnamefont {Sørensen}},\ and\
  \bibinfo {author} {\bibfnamefont {M.~D.}\ \bibnamefont {Lukin}},\ }\bibfield
  {title} {\bibinfo {title} {Distributed quantum computation based on small
  quantum registers},\ }\href {https://doi.org/10.1103/physreva.76.062323}
  {\bibfield  {journal} {\bibinfo  {journal} {Physical Review A}\ }\textbf
  {\bibinfo {volume} {76}},\ \bibinfo {pages} {062323} (\bibinfo {year}
  {2007})}\BibitemShut {NoStop}%
\bibitem [{\citenamefont {Monroe}\ \emph {et~al.}(2014)\citenamefont {Monroe},
  \citenamefont {Raussendorf}, \citenamefont {Ruthven}, \citenamefont {Brown},
  \citenamefont {Maunz}, \citenamefont {Duan},\ and\ \citenamefont
  {Kim}}]{Monroe2014}%
  \BibitemOpen
  \bibfield  {author} {\bibinfo {author} {\bibfnamefont {C.}~\bibnamefont
  {Monroe}}, \bibinfo {author} {\bibfnamefont {R.}~\bibnamefont {Raussendorf}},
  \bibinfo {author} {\bibfnamefont {A.}~\bibnamefont {Ruthven}}, \bibinfo
  {author} {\bibfnamefont {K.~R.}\ \bibnamefont {Brown}}, \bibinfo {author}
  {\bibfnamefont {P.}~\bibnamefont {Maunz}}, \bibinfo {author} {\bibfnamefont
  {L.-M.}\ \bibnamefont {Duan}},\ and\ \bibinfo {author} {\bibfnamefont
  {J.}~\bibnamefont {Kim}},\ }\bibfield  {title} {\bibinfo {title} {Large-scale
  modular quantum-computer architecture with atomic memory and photonic
  interconnects},\ }\href {https://doi.org/10.1103/physreva.89.022317}
  {\bibfield  {journal} {\bibinfo  {journal} {Physical Review A}\ }\textbf
  {\bibinfo {volume} {89}},\ \bibinfo {pages} {022317} (\bibinfo {year}
  {2014})}\BibitemShut {NoStop}%
\bibitem [{\citenamefont {Proctor}\ \emph {et~al.}(2018)\citenamefont
  {Proctor}, \citenamefont {Knott},\ and\ \citenamefont
  {Dunningham}}]{Proctor2018}%
  \BibitemOpen
  \bibfield  {author} {\bibinfo {author} {\bibfnamefont {T.~J.}\ \bibnamefont
  {Proctor}}, \bibinfo {author} {\bibfnamefont {P.~A.}\ \bibnamefont {Knott}},\
  and\ \bibinfo {author} {\bibfnamefont {J.~A.}\ \bibnamefont {Dunningham}},\
  }\bibfield  {title} {\bibinfo {title} {Multiparameter estimation in networked
  quantum sensors},\ }\href {https://doi.org/10.1103/PhysRevLett.120.080501}
  {\bibfield  {journal} {\bibinfo  {journal} {Physical Review Letters}\
  }\textbf {\bibinfo {volume} {120}},\ \bibinfo {pages} {080501} (\bibinfo
  {year} {2018})}\BibitemShut {NoStop}%
\bibitem [{\citenamefont {Zhang}\ and\ \citenamefont
  {Zhuang}(2021)}]{Zhang2021}%
  \BibitemOpen
  \bibfield  {author} {\bibinfo {author} {\bibfnamefont {Z.}~\bibnamefont
  {Zhang}}\ and\ \bibinfo {author} {\bibfnamefont {Q.}~\bibnamefont {Zhuang}},\
  }\bibfield  {title} {\bibinfo {title} {Distributed quantum sensing},\ }\href
  {https://doi.org/10.1088/2058-9565/abd4c3} {\bibfield  {journal} {\bibinfo
  {journal} {Quantum Science and Technology}\ }\textbf {\bibinfo {volume}
  {6}},\ \bibinfo {pages} {043001} (\bibinfo {year} {2021})}\BibitemShut
  {NoStop}%
\bibitem [{\citenamefont {Bravyi}\ \emph {et~al.}(2022)\citenamefont {Bravyi},
  \citenamefont {Dial}, \citenamefont {Gambetta}, \citenamefont {Gil},\ and\
  \citenamefont {Nazario}}]{Bravyi2022}%
  \BibitemOpen
  \bibfield  {author} {\bibinfo {author} {\bibfnamefont {S.}~\bibnamefont
  {Bravyi}}, \bibinfo {author} {\bibfnamefont {O.}~\bibnamefont {Dial}},
  \bibinfo {author} {\bibfnamefont {J.~M.}\ \bibnamefont {Gambetta}}, \bibinfo
  {author} {\bibfnamefont {D.}~\bibnamefont {Gil}},\ and\ \bibinfo {author}
  {\bibfnamefont {Z.}~\bibnamefont {Nazario}},\ }\bibfield  {title} {\bibinfo
  {title} {The future of quantum computing with superconducting qubits},\
  }\href {https://doi.org/10.1063/5.0082975} {\bibfield  {journal} {\bibinfo
  {journal} {Journal of Applied Physics}\ }\textbf {\bibinfo {volume} {132}},\
  \bibinfo {pages} {160902} (\bibinfo {year} {2022})}\BibitemShut {NoStop}%
\bibitem [{\citenamefont {Axline}\ \emph {et~al.}(2018)\citenamefont {Axline},
  \citenamefont {Burkhart}, \citenamefont {Pfaff}, \citenamefont {Zhang},
  \citenamefont {Chou}, \citenamefont {Campagne-Ibarcq}, \citenamefont
  {Reinhold}, \citenamefont {Frunzio}, \citenamefont {Girvin}, \citenamefont
  {Jiang}, \citenamefont {Devoret},\ and\ \citenamefont
  {Schoelkopf}}]{Axline2018}%
  \BibitemOpen
  \bibfield  {author} {\bibinfo {author} {\bibfnamefont {C.~J.}\ \bibnamefont
  {Axline}}, \bibinfo {author} {\bibfnamefont {L.~D.}\ \bibnamefont
  {Burkhart}}, \bibinfo {author} {\bibfnamefont {W.}~\bibnamefont {Pfaff}},
  \bibinfo {author} {\bibfnamefont {M.}~\bibnamefont {Zhang}}, \bibinfo
  {author} {\bibfnamefont {K.}~\bibnamefont {Chou}}, \bibinfo {author}
  {\bibfnamefont {P.}~\bibnamefont {Campagne-Ibarcq}}, \bibinfo {author}
  {\bibfnamefont {P.}~\bibnamefont {Reinhold}}, \bibinfo {author}
  {\bibfnamefont {L.}~\bibnamefont {Frunzio}}, \bibinfo {author} {\bibfnamefont
  {S.~M.}\ \bibnamefont {Girvin}}, \bibinfo {author} {\bibfnamefont
  {L.}~\bibnamefont {Jiang}}, \bibinfo {author} {\bibfnamefont {M.~H.}\
  \bibnamefont {Devoret}},\ and\ \bibinfo {author} {\bibfnamefont {R.~J.}\
  \bibnamefont {Schoelkopf}},\ }\bibfield  {title} {\bibinfo {title} {On-demand
  quantum state transfer and entanglement between remote microwave cavity
  memories},\ }\href {https://doi.org/10.1038/s41567-018-0115-y} {\bibfield
  {journal} {\bibinfo  {journal} {Nature Physics}\ }\textbf {\bibinfo {volume}
  {14}},\ \bibinfo {pages} {705} (\bibinfo {year} {2018})}\BibitemShut
  {NoStop}%
\bibitem [{\citenamefont {Kurpiers}\ \emph {et~al.}(2018)\citenamefont
  {Kurpiers}, \citenamefont {Magnard}, \citenamefont {Walter}, \citenamefont
  {Royer}, \citenamefont {Pechal}, \citenamefont {Heinsoo}, \citenamefont
  {Salathé}, \citenamefont {Akin}, \citenamefont {Storz}, \citenamefont
  {Besse}, \citenamefont {Gasparinetti}, \citenamefont {Blais},\ and\
  \citenamefont {Wallraff}}]{Kurpiers2018}%
  \BibitemOpen
  \bibfield  {author} {\bibinfo {author} {\bibfnamefont {P.}~\bibnamefont
  {Kurpiers}}, \bibinfo {author} {\bibfnamefont {P.}~\bibnamefont {Magnard}},
  \bibinfo {author} {\bibfnamefont {T.}~\bibnamefont {Walter}}, \bibinfo
  {author} {\bibfnamefont {B.}~\bibnamefont {Royer}}, \bibinfo {author}
  {\bibfnamefont {M.}~\bibnamefont {Pechal}}, \bibinfo {author} {\bibfnamefont
  {J.}~\bibnamefont {Heinsoo}}, \bibinfo {author} {\bibfnamefont
  {Y.}~\bibnamefont {Salathé}}, \bibinfo {author} {\bibfnamefont
  {A.}~\bibnamefont {Akin}}, \bibinfo {author} {\bibfnamefont {S.}~\bibnamefont
  {Storz}}, \bibinfo {author} {\bibfnamefont {J.-C.}\ \bibnamefont {Besse}},
  \bibinfo {author} {\bibfnamefont {S.}~\bibnamefont {Gasparinetti}}, \bibinfo
  {author} {\bibfnamefont {A.}~\bibnamefont {Blais}},\ and\ \bibinfo {author}
  {\bibfnamefont {A.}~\bibnamefont {Wallraff}},\ }\bibfield  {title} {\bibinfo
  {title} {Deterministic quantum state transfer and remote entanglement using
  microwave photons},\ }\href {https://doi.org/10.1038/s41586-018-0195-y}
  {\bibfield  {journal} {\bibinfo  {journal} {Nature}\ }\textbf {\bibinfo
  {volume} {558}},\ \bibinfo {pages} {264} (\bibinfo {year}
  {2018})}\BibitemShut {NoStop}%
\bibitem [{\citenamefont {Campagne-Ibarcq}\ \emph {et~al.}(2018)\citenamefont
  {Campagne-Ibarcq}, \citenamefont {Zalys-Geller}, \citenamefont {Narla},
  \citenamefont {Shankar}, \citenamefont {Reinhold}, \citenamefont {Burkhart},
  \citenamefont {Axline}, \citenamefont {Pfaff}, \citenamefont {Frunzio},
  \citenamefont {Schoelkopf},\ and\ \citenamefont
  {Devoret}}]{CampagneIbarcq2018}%
  \BibitemOpen
  \bibfield  {author} {\bibinfo {author} {\bibfnamefont {P.}~\bibnamefont
  {Campagne-Ibarcq}}, \bibinfo {author} {\bibfnamefont {E.}~\bibnamefont
  {Zalys-Geller}}, \bibinfo {author} {\bibfnamefont {A.}~\bibnamefont {Narla}},
  \bibinfo {author} {\bibfnamefont {S.}~\bibnamefont {Shankar}}, \bibinfo
  {author} {\bibfnamefont {P.}~\bibnamefont {Reinhold}}, \bibinfo {author}
  {\bibfnamefont {L.}~\bibnamefont {Burkhart}}, \bibinfo {author}
  {\bibfnamefont {C.}~\bibnamefont {Axline}}, \bibinfo {author} {\bibfnamefont
  {W.}~\bibnamefont {Pfaff}}, \bibinfo {author} {\bibfnamefont
  {L.}~\bibnamefont {Frunzio}}, \bibinfo {author} {\bibfnamefont
  {R.}~\bibnamefont {Schoelkopf}},\ and\ \bibinfo {author} {\bibfnamefont
  {M.}~\bibnamefont {Devoret}},\ }\bibfield  {title} {\bibinfo {title}
  {Deterministic remote entanglement of superconducting circuits through
  microwave two-photon transitions},\ }\href
  {https://doi.org/10.1103/physrevlett.120.200501} {\bibfield  {journal}
  {\bibinfo  {journal} {Physical Review Letters}\ }\textbf {\bibinfo {volume}
  {120}},\ \bibinfo {pages} {200501} (\bibinfo {year} {2018})}\BibitemShut
  {NoStop}%
\bibitem [{\citenamefont {Leung}\ \emph {et~al.}(2019)\citenamefont {Leung},
  \citenamefont {Lu}, \citenamefont {Chakram}, \citenamefont {Naik},
  \citenamefont {Earnest}, \citenamefont {Ma}, \citenamefont {Jacobs},
  \citenamefont {Cleland},\ and\ \citenamefont {Schuster}}]{Leung2019}%
  \BibitemOpen
  \bibfield  {author} {\bibinfo {author} {\bibfnamefont {N.}~\bibnamefont
  {Leung}}, \bibinfo {author} {\bibfnamefont {Y.}~\bibnamefont {Lu}}, \bibinfo
  {author} {\bibfnamefont {S.}~\bibnamefont {Chakram}}, \bibinfo {author}
  {\bibfnamefont {R.~K.}\ \bibnamefont {Naik}}, \bibinfo {author}
  {\bibfnamefont {N.}~\bibnamefont {Earnest}}, \bibinfo {author} {\bibfnamefont
  {R.}~\bibnamefont {Ma}}, \bibinfo {author} {\bibfnamefont {K.}~\bibnamefont
  {Jacobs}}, \bibinfo {author} {\bibfnamefont {A.~N.}\ \bibnamefont
  {Cleland}},\ and\ \bibinfo {author} {\bibfnamefont {D.~I.}\ \bibnamefont
  {Schuster}},\ }\bibfield  {title} {\bibinfo {title} {Deterministic
  bidirectional communication and remote entanglement generation between
  superconducting qubits},\ }\href {https://doi.org/10.1038/s41534-019-0128-0}
  {\bibfield  {journal} {\bibinfo  {journal} {npj Quantum Information}\
  }\textbf {\bibinfo {volume} {5}},\ \bibinfo {pages} {18} (\bibinfo {year}
  {2019})}\BibitemShut {NoStop}%
\bibitem [{\citenamefont {Zhong}\ \emph {et~al.}(2019)\citenamefont {Zhong},
  \citenamefont {Chang}, \citenamefont {Satzinger}, \citenamefont {Chou},
  \citenamefont {Bienfait}, \citenamefont {Conner}, \citenamefont {Dumur},
  \citenamefont {Grebel}, \citenamefont {Peairs}, \citenamefont {Povey},
  \citenamefont {Schuster},\ and\ \citenamefont {Cleland}}]{Zhong2019}%
  \BibitemOpen
  \bibfield  {author} {\bibinfo {author} {\bibfnamefont {Y.~P.}\ \bibnamefont
  {Zhong}}, \bibinfo {author} {\bibfnamefont {H.-S.}\ \bibnamefont {Chang}},
  \bibinfo {author} {\bibfnamefont {K.~J.}\ \bibnamefont {Satzinger}}, \bibinfo
  {author} {\bibfnamefont {M.-H.}\ \bibnamefont {Chou}}, \bibinfo {author}
  {\bibfnamefont {A.}~\bibnamefont {Bienfait}}, \bibinfo {author}
  {\bibfnamefont {C.~R.}\ \bibnamefont {Conner}}, \bibinfo {author}
  {\bibfnamefont {Ã.}~\bibnamefont {Dumur}}, \bibinfo {author} {\bibfnamefont
  {J.}~\bibnamefont {Grebel}}, \bibinfo {author} {\bibfnamefont {G.~A.}\
  \bibnamefont {Peairs}}, \bibinfo {author} {\bibfnamefont {R.~G.}\
  \bibnamefont {Povey}}, \bibinfo {author} {\bibfnamefont {D.~I.}\ \bibnamefont
  {Schuster}},\ and\ \bibinfo {author} {\bibfnamefont {A.~N.}\ \bibnamefont
  {Cleland}},\ }\bibfield  {title} {\bibinfo {title} {Violating {Bell's}
  inequality with remotely connected superconducting qubits},\ }\href
  {https://doi.org/10.1038/s41567-019-0507-7} {\bibfield  {journal} {\bibinfo
  {journal} {Nature Physics}\ }\textbf {\bibinfo {volume} {15}},\ \bibinfo
  {pages} {741} (\bibinfo {year} {2019})}\BibitemShut {NoStop}%
\bibitem [{\citenamefont {Chang}\ \emph {et~al.}(2020)\citenamefont {Chang},
  \citenamefont {Zhong}, \citenamefont {Bienfait}, \citenamefont {Chou},
  \citenamefont {Conner}, \citenamefont {Dumur}, \citenamefont {Grebel},
  \citenamefont {Peairs}, \citenamefont {Povey}, \citenamefont {Satzinger},\
  and\ \citenamefont {Cleland}}]{Chang2020}%
  \BibitemOpen
  \bibfield  {author} {\bibinfo {author} {\bibfnamefont {H.-S.}\ \bibnamefont
  {Chang}}, \bibinfo {author} {\bibfnamefont {Y.}~\bibnamefont {Zhong}},
  \bibinfo {author} {\bibfnamefont {A.}~\bibnamefont {Bienfait}}, \bibinfo
  {author} {\bibfnamefont {M.-H.}\ \bibnamefont {Chou}}, \bibinfo {author}
  {\bibfnamefont {C.}~\bibnamefont {Conner}}, \bibinfo {author} {\bibfnamefont
  {Ã.}~\bibnamefont {Dumur}}, \bibinfo {author} {\bibfnamefont
  {J.}~\bibnamefont {Grebel}}, \bibinfo {author} {\bibfnamefont
  {G.}~\bibnamefont {Peairs}}, \bibinfo {author} {\bibfnamefont
  {R.}~\bibnamefont {Povey}}, \bibinfo {author} {\bibfnamefont
  {K.}~\bibnamefont {Satzinger}},\ and\ \bibinfo {author} {\bibfnamefont
  {A.}~\bibnamefont {Cleland}},\ }\bibfield  {title} {\bibinfo {title} {Remote
  entanglement via adiabatic passage using a tunably dissipative quantum
  communication system},\ }\href
  {https://doi.org/10.1103/physrevlett.124.240502} {\bibfield  {journal}
  {\bibinfo  {journal} {Physical Review Letters}\ }\textbf {\bibinfo {volume}
  {124}},\ \bibinfo {pages} {240502} (\bibinfo {year} {2020})}\BibitemShut
  {NoStop}%
\bibitem [{\citenamefont {Burkhart}\ \emph {et~al.}(2021)\citenamefont
  {Burkhart}, \citenamefont {Teoh}, \citenamefont {Zhang}, \citenamefont
  {Axline}, \citenamefont {Frunzio}, \citenamefont {Devoret}, \citenamefont
  {Jiang}, \citenamefont {Girvin},\ and\ \citenamefont
  {Schoelkopf}}]{Burkhart2021}%
  \BibitemOpen
  \bibfield  {author} {\bibinfo {author} {\bibfnamefont {L.~D.}\ \bibnamefont
  {Burkhart}}, \bibinfo {author} {\bibfnamefont {J.~D.}\ \bibnamefont {Teoh}},
  \bibinfo {author} {\bibfnamefont {Y.}~\bibnamefont {Zhang}}, \bibinfo
  {author} {\bibfnamefont {C.~J.}\ \bibnamefont {Axline}}, \bibinfo {author}
  {\bibfnamefont {L.}~\bibnamefont {Frunzio}}, \bibinfo {author} {\bibfnamefont
  {M.}~\bibnamefont {Devoret}}, \bibinfo {author} {\bibfnamefont
  {L.}~\bibnamefont {Jiang}}, \bibinfo {author} {\bibfnamefont
  {S.}~\bibnamefont {Girvin}},\ and\ \bibinfo {author} {\bibfnamefont
  {R.}~\bibnamefont {Schoelkopf}},\ }\bibfield  {title} {\bibinfo {title}
  {Error-detected state transfer and entanglement in a superconducting quantum
  network},\ }\href {https://doi.org/10.1103/prxquantum.2.030321} {\bibfield
  {journal} {\bibinfo  {journal} {PRX Quantum}\ }\textbf {\bibinfo {volume}
  {2}},\ \bibinfo {pages} {030321} (\bibinfo {year} {2021})}\BibitemShut
  {NoStop}%
\bibitem [{\citenamefont {Zhong}\ \emph {et~al.}(2021)\citenamefont {Zhong},
  \citenamefont {Chang}, \citenamefont {Bienfait}, \citenamefont {Dumur},
  \citenamefont {Chou}, \citenamefont {Conner}, \citenamefont {Grebel},
  \citenamefont {Povey}, \citenamefont {Yan}, \citenamefont {Schuster},\ and\
  \citenamefont {Cleland}}]{Zhong2021}%
  \BibitemOpen
  \bibfield  {author} {\bibinfo {author} {\bibfnamefont {Y.}~\bibnamefont
  {Zhong}}, \bibinfo {author} {\bibfnamefont {H.-S.}\ \bibnamefont {Chang}},
  \bibinfo {author} {\bibfnamefont {A.}~\bibnamefont {Bienfait}}, \bibinfo
  {author} {\bibfnamefont {Ã.}~\bibnamefont {Dumur}}, \bibinfo {author}
  {\bibfnamefont {M.-H.}\ \bibnamefont {Chou}}, \bibinfo {author}
  {\bibfnamefont {C.~R.}\ \bibnamefont {Conner}}, \bibinfo {author}
  {\bibfnamefont {J.}~\bibnamefont {Grebel}}, \bibinfo {author} {\bibfnamefont
  {R.~G.}\ \bibnamefont {Povey}}, \bibinfo {author} {\bibfnamefont
  {H.}~\bibnamefont {Yan}}, \bibinfo {author} {\bibfnamefont {D.~I.}\
  \bibnamefont {Schuster}},\ and\ \bibinfo {author} {\bibfnamefont {A.~N.}\
  \bibnamefont {Cleland}},\ }\bibfield  {title} {\bibinfo {title}
  {Deterministic multi-qubit entanglement in a quantum network},\ }\href
  {https://doi.org/10.1038/s41586-021-03288-7} {\bibfield  {journal} {\bibinfo
  {journal} {Nature}\ }\textbf {\bibinfo {volume} {590}},\ \bibinfo {pages}
  {571} (\bibinfo {year} {2021})}\BibitemShut {NoStop}%
\bibitem [{\citenamefont {Storz}\ \emph {et~al.}(2023)\citenamefont {Storz},
  \citenamefont {Schär}, \citenamefont {Kulikov}, \citenamefont {Magnard},
  \citenamefont {Kurpiers}, \citenamefont {Lütolf}, \citenamefont {Walter},
  \citenamefont {Copetudo}, \citenamefont {Reuer}, \citenamefont {Akin},
  \citenamefont {Besse}, \citenamefont {Gabureac}, \citenamefont {Norris},
  \citenamefont {Rosario}, \citenamefont {Martin}, \citenamefont {Martinez},
  \citenamefont {Amaya}, \citenamefont {Mitchell}, \citenamefont {Abellan},
  \citenamefont {Bancal}, \citenamefont {Sangouard}, \citenamefont {Royer},
  \citenamefont {Blais},\ and\ \citenamefont {Wallraff}}]{Storz2023}%
  \BibitemOpen
  \bibfield  {author} {\bibinfo {author} {\bibfnamefont {S.}~\bibnamefont
  {Storz}}, \bibinfo {author} {\bibfnamefont {J.}~\bibnamefont {Schär}},
  \bibinfo {author} {\bibfnamefont {A.}~\bibnamefont {Kulikov}}, \bibinfo
  {author} {\bibfnamefont {P.}~\bibnamefont {Magnard}}, \bibinfo {author}
  {\bibfnamefont {P.}~\bibnamefont {Kurpiers}}, \bibinfo {author}
  {\bibfnamefont {J.}~\bibnamefont {Lütolf}}, \bibinfo {author} {\bibfnamefont
  {T.}~\bibnamefont {Walter}}, \bibinfo {author} {\bibfnamefont
  {A.}~\bibnamefont {Copetudo}}, \bibinfo {author} {\bibfnamefont
  {K.}~\bibnamefont {Reuer}}, \bibinfo {author} {\bibfnamefont
  {A.}~\bibnamefont {Akin}}, \bibinfo {author} {\bibfnamefont {J.-C.}\
  \bibnamefont {Besse}}, \bibinfo {author} {\bibfnamefont {M.}~\bibnamefont
  {Gabureac}}, \bibinfo {author} {\bibfnamefont {G.~J.}\ \bibnamefont
  {Norris}}, \bibinfo {author} {\bibfnamefont {A.}~\bibnamefont {Rosario}},
  \bibinfo {author} {\bibfnamefont {F.}~\bibnamefont {Martin}}, \bibinfo
  {author} {\bibfnamefont {J.}~\bibnamefont {Martinez}}, \bibinfo {author}
  {\bibfnamefont {W.}~\bibnamefont {Amaya}}, \bibinfo {author} {\bibfnamefont
  {M.~W.}\ \bibnamefont {Mitchell}}, \bibinfo {author} {\bibfnamefont
  {C.}~\bibnamefont {Abellan}}, \bibinfo {author} {\bibfnamefont {J.-D.}\
  \bibnamefont {Bancal}}, \bibinfo {author} {\bibfnamefont {N.}~\bibnamefont
  {Sangouard}}, \bibinfo {author} {\bibfnamefont {B.}~\bibnamefont {Royer}},
  \bibinfo {author} {\bibfnamefont {A.}~\bibnamefont {Blais}},\ and\ \bibinfo
  {author} {\bibfnamefont {A.}~\bibnamefont {Wallraff}},\ }\bibfield  {title}
  {\bibinfo {title} {Loophole-free {Bell} inequality violation with
  superconducting circuits},\ }\href
  {https://doi.org/10.1038/s41586-023-05885-0} {\bibfield  {journal} {\bibinfo
  {journal} {Nature}\ }\textbf {\bibinfo {volume} {617}},\ \bibinfo {pages}
  {265} (\bibinfo {year} {2023})}\BibitemShut {NoStop}%
\bibitem [{\citenamefont {Grebel}\ \emph {et~al.}(2024)\citenamefont {Grebel},
  \citenamefont {Yan}, \citenamefont {Chou}, \citenamefont {Andersson},
  \citenamefont {Conner}, \citenamefont {Joshi}, \citenamefont {Miller},
  \citenamefont {Povey}, \citenamefont {Qiao}, \citenamefont {Wu},\ and\
  \citenamefont {Cleland}}]{Grebel2024}%
  \BibitemOpen
  \bibfield  {author} {\bibinfo {author} {\bibfnamefont {J.}~\bibnamefont
  {Grebel}}, \bibinfo {author} {\bibfnamefont {H.}~\bibnamefont {Yan}},
  \bibinfo {author} {\bibfnamefont {M.-H.}\ \bibnamefont {Chou}}, \bibinfo
  {author} {\bibfnamefont {G.}~\bibnamefont {Andersson}}, \bibinfo {author}
  {\bibfnamefont {C.~R.}\ \bibnamefont {Conner}}, \bibinfo {author}
  {\bibfnamefont {Y.~J.}\ \bibnamefont {Joshi}}, \bibinfo {author}
  {\bibfnamefont {J.~M.}\ \bibnamefont {Miller}}, \bibinfo {author}
  {\bibfnamefont {R.~G.}\ \bibnamefont {Povey}}, \bibinfo {author}
  {\bibfnamefont {H.}~\bibnamefont {Qiao}}, \bibinfo {author} {\bibfnamefont
  {X.}~\bibnamefont {Wu}},\ and\ \bibinfo {author} {\bibfnamefont {A.~N.}\
  \bibnamefont {Cleland}},\ }\bibfield  {title} {\bibinfo {title}
  {Bidirectional multiphoton communication between remote superconducting
  nodes},\ }\href {https://doi.org/10.1103/physrevlett.132.047001} {\bibfield
  {journal} {\bibinfo  {journal} {Physical Review Letters}\ }\textbf {\bibinfo
  {volume} {132}},\ \bibinfo {pages} {047001} (\bibinfo {year}
  {2024})}\BibitemShut {NoStop}%
\bibitem [{\citenamefont {Niu}\ \emph {et~al.}(2023)\citenamefont {Niu},
  \citenamefont {Zhang}, \citenamefont {Liu}, \citenamefont {Qiu},
  \citenamefont {Huang}, \citenamefont {Huang}, \citenamefont {Jia},
  \citenamefont {Liu}, \citenamefont {Tao}, \citenamefont {Wei}, \citenamefont
  {Zhou}, \citenamefont {Zou}, \citenamefont {Chen}, \citenamefont {Deng},
  \citenamefont {Deng}, \citenamefont {Hu}, \citenamefont {Hu}, \citenamefont
  {Li}, \citenamefont {Tan}, \citenamefont {Xu}, \citenamefont {Yan},
  \citenamefont {Yan}, \citenamefont {Liu}, \citenamefont {Zhong},
  \citenamefont {Cleland},\ and\ \citenamefont {Yu}}]{Niu2023}%
  \BibitemOpen
  \bibfield  {author} {\bibinfo {author} {\bibfnamefont {J.}~\bibnamefont
  {Niu}}, \bibinfo {author} {\bibfnamefont {L.}~\bibnamefont {Zhang}}, \bibinfo
  {author} {\bibfnamefont {Y.}~\bibnamefont {Liu}}, \bibinfo {author}
  {\bibfnamefont {J.}~\bibnamefont {Qiu}}, \bibinfo {author} {\bibfnamefont
  {W.}~\bibnamefont {Huang}}, \bibinfo {author} {\bibfnamefont
  {J.}~\bibnamefont {Huang}}, \bibinfo {author} {\bibfnamefont
  {H.}~\bibnamefont {Jia}}, \bibinfo {author} {\bibfnamefont {J.}~\bibnamefont
  {Liu}}, \bibinfo {author} {\bibfnamefont {Z.}~\bibnamefont {Tao}}, \bibinfo
  {author} {\bibfnamefont {W.}~\bibnamefont {Wei}}, \bibinfo {author}
  {\bibfnamefont {Y.}~\bibnamefont {Zhou}}, \bibinfo {author} {\bibfnamefont
  {W.}~\bibnamefont {Zou}}, \bibinfo {author} {\bibfnamefont {Y.}~\bibnamefont
  {Chen}}, \bibinfo {author} {\bibfnamefont {X.}~\bibnamefont {Deng}}, \bibinfo
  {author} {\bibfnamefont {X.}~\bibnamefont {Deng}}, \bibinfo {author}
  {\bibfnamefont {C.}~\bibnamefont {Hu}}, \bibinfo {author} {\bibfnamefont
  {L.}~\bibnamefont {Hu}}, \bibinfo {author} {\bibfnamefont {J.}~\bibnamefont
  {Li}}, \bibinfo {author} {\bibfnamefont {D.}~\bibnamefont {Tan}}, \bibinfo
  {author} {\bibfnamefont {Y.}~\bibnamefont {Xu}}, \bibinfo {author}
  {\bibfnamefont {F.}~\bibnamefont {Yan}}, \bibinfo {author} {\bibfnamefont
  {T.}~\bibnamefont {Yan}}, \bibinfo {author} {\bibfnamefont {S.}~\bibnamefont
  {Liu}}, \bibinfo {author} {\bibfnamefont {Y.}~\bibnamefont {Zhong}}, \bibinfo
  {author} {\bibfnamefont {A.~N.}\ \bibnamefont {Cleland}},\ and\ \bibinfo
  {author} {\bibfnamefont {D.}~\bibnamefont {Yu}},\ }\bibfield  {title}
  {\bibinfo {title} {Low-loss interconnects for modular superconducting quantum
  processors},\ }\href {https://doi.org/10.1038/s41928-023-00925-z} {\bibfield
  {journal} {\bibinfo  {journal} {Nature Electronics}\ }\textbf {\bibinfo
  {volume} {6}},\ \bibinfo {pages} {235} (\bibinfo {year} {2023})}\BibitemShut
  {NoStop}%
\bibitem [{\citenamefont {Yan}\ \emph {et~al.}(2022)\citenamefont {Yan},
  \citenamefont {Zhong}, \citenamefont {Chang}, \citenamefont {Bienfait},
  \citenamefont {Chou}, \citenamefont {Conner}, \citenamefont {Dumur},
  \citenamefont {Grebel}, \citenamefont {Povey},\ and\ \citenamefont
  {Cleland}}]{Yan2022}%
  \BibitemOpen
  \bibfield  {author} {\bibinfo {author} {\bibfnamefont {H.}~\bibnamefont
  {Yan}}, \bibinfo {author} {\bibfnamefont {Y.}~\bibnamefont {Zhong}}, \bibinfo
  {author} {\bibfnamefont {H.-S.}\ \bibnamefont {Chang}}, \bibinfo {author}
  {\bibfnamefont {A.}~\bibnamefont {Bienfait}}, \bibinfo {author}
  {\bibfnamefont {M.-H.}\ \bibnamefont {Chou}}, \bibinfo {author}
  {\bibfnamefont {C.~R.}\ \bibnamefont {Conner}}, \bibinfo {author}
  {\bibfnamefont {Ã.}~\bibnamefont {Dumur}}, \bibinfo {author} {\bibfnamefont
  {J.}~\bibnamefont {Grebel}}, \bibinfo {author} {\bibfnamefont {R.~G.}\
  \bibnamefont {Povey}},\ and\ \bibinfo {author} {\bibfnamefont {A.~N.}\
  \bibnamefont {Cleland}},\ }\bibfield  {title} {\bibinfo {title} {Entanglement
  purification and protection in a superconducting quantum network},\ }\href
  {https://doi.org/10.1103/physrevlett.128.080504} {\bibfield  {journal}
  {\bibinfo  {journal} {Physical Review Letters}\ }\textbf {\bibinfo {volume}
  {128}},\ \bibinfo {pages} {080504} (\bibinfo {year} {2022})}\BibitemShut
  {NoStop}%
\bibitem [{\citenamefont {Chou}\ \emph {et~al.}(2018)\citenamefont {Chou},
  \citenamefont {Blumoff}, \citenamefont {Wang}, \citenamefont {Reinhold},
  \citenamefont {Axline}, \citenamefont {Gao}, \citenamefont {Frunzio},
  \citenamefont {Devoret}, \citenamefont {Jiang},\ and\ \citenamefont
  {Schoelkopf}}]{Chou2018}%
  \BibitemOpen
  \bibfield  {author} {\bibinfo {author} {\bibfnamefont {K.~S.}\ \bibnamefont
  {Chou}}, \bibinfo {author} {\bibfnamefont {J.~Z.}\ \bibnamefont {Blumoff}},
  \bibinfo {author} {\bibfnamefont {C.~S.}\ \bibnamefont {Wang}}, \bibinfo
  {author} {\bibfnamefont {P.~C.}\ \bibnamefont {Reinhold}}, \bibinfo {author}
  {\bibfnamefont {C.~J.}\ \bibnamefont {Axline}}, \bibinfo {author}
  {\bibfnamefont {Y.~Y.}\ \bibnamefont {Gao}}, \bibinfo {author} {\bibfnamefont
  {L.}~\bibnamefont {Frunzio}}, \bibinfo {author} {\bibfnamefont {M.~H.}\
  \bibnamefont {Devoret}}, \bibinfo {author} {\bibfnamefont {L.}~\bibnamefont
  {Jiang}},\ and\ \bibinfo {author} {\bibfnamefont {R.~J.}\ \bibnamefont
  {Schoelkopf}},\ }\bibfield  {title} {\bibinfo {title} {Deterministic
  teleportation of a quantum gate between two logical qubits},\ }\href
  {https://doi.org/10.1038/s41586-018-0470-y} {\bibfield  {journal} {\bibinfo
  {journal} {Nature}\ }\textbf {\bibinfo {volume} {561}},\ \bibinfo {pages}
  {368} (\bibinfo {year} {2018})}\BibitemShut {NoStop}%
\bibitem [{\citenamefont {Qiu}\ \emph {et~al.}(2025)\citenamefont {Qiu},
  \citenamefont {Liu}, \citenamefont {Hu}, \citenamefont {Wu}, \citenamefont
  {Niu}, \citenamefont {Zhang}, \citenamefont {Huang}, \citenamefont {Chen},
  \citenamefont {Li}, \citenamefont {Liu}, \citenamefont {Zhong}, \citenamefont
  {Duan},\ and\ \citenamefont {Yu}}]{Qiu2025}%
  \BibitemOpen
  \bibfield  {author} {\bibinfo {author} {\bibfnamefont {J.}~\bibnamefont
  {Qiu}}, \bibinfo {author} {\bibfnamefont {Y.}~\bibnamefont {Liu}}, \bibinfo
  {author} {\bibfnamefont {L.}~\bibnamefont {Hu}}, \bibinfo {author}
  {\bibfnamefont {Y.}~\bibnamefont {Wu}}, \bibinfo {author} {\bibfnamefont
  {J.}~\bibnamefont {Niu}}, \bibinfo {author} {\bibfnamefont {L.}~\bibnamefont
  {Zhang}}, \bibinfo {author} {\bibfnamefont {W.}~\bibnamefont {Huang}},
  \bibinfo {author} {\bibfnamefont {Y.}~\bibnamefont {Chen}}, \bibinfo {author}
  {\bibfnamefont {J.}~\bibnamefont {Li}}, \bibinfo {author} {\bibfnamefont
  {S.}~\bibnamefont {Liu}}, \bibinfo {author} {\bibfnamefont {Y.}~\bibnamefont
  {Zhong}}, \bibinfo {author} {\bibfnamefont {L.}~\bibnamefont {Duan}},\ and\
  \bibinfo {author} {\bibfnamefont {D.}~\bibnamefont {Yu}},\ }\bibfield
  {title} {\bibinfo {title} {Deterministic quantum state and gate teleportation
  between distant superconducting chips},\ }\href
  {https://doi.org/10.1016/j.scib.2024.11.047} {\bibfield  {journal} {\bibinfo
  {journal} {Science Bulletin}\ }\textbf {\bibinfo {volume} {70}},\ \bibinfo
  {pages} {351–358} (\bibinfo {year} {2025})}\BibitemShut {NoStop}%
\bibitem [{\citenamefont {{\.Z}ukowski}\ \emph {et~al.}(1993)\citenamefont
  {{\.Z}ukowski}, \citenamefont {Zeilinger}, \citenamefont {Horne},\ and\
  \citenamefont {Ekert}}]{Zukowski1993}%
  \BibitemOpen
  \bibfield  {author} {\bibinfo {author} {\bibfnamefont {M.}~\bibnamefont
  {{\.Z}ukowski}}, \bibinfo {author} {\bibfnamefont {A.}~\bibnamefont
  {Zeilinger}}, \bibinfo {author} {\bibfnamefont {M.~A.}\ \bibnamefont
  {Horne}},\ and\ \bibinfo {author} {\bibfnamefont {A.~K.}\ \bibnamefont
  {Ekert}},\ }\bibfield  {title} {\bibinfo {title} {``{E}vent-ready-detectors''
  {B}ell experiment via entanglement swapping},\ }\href
  {https://doi.org/10.1103/physrevlett.71.4287} {\bibfield  {journal} {\bibinfo
   {journal} {Physical Review Letters}\ }\textbf {\bibinfo {volume} {71}},\
  \bibinfo {pages} {4287} (\bibinfo {year} {1993})}\BibitemShut {NoStop}%
\bibitem [{\citenamefont {G{\"u}hne}\ and\ \citenamefont
  {Seevinck}(2010)}]{Guehne2010}%
  \BibitemOpen
  \bibfield  {author} {\bibinfo {author} {\bibfnamefont {O.}~\bibnamefont
  {G{\"u}hne}}\ and\ \bibinfo {author} {\bibfnamefont {M.}~\bibnamefont
  {Seevinck}},\ }\bibfield  {title} {\bibinfo {title} {Separability criteria
  for genuine multiparticle entanglement},\ }\href
  {https://doi.org/10.1088/1367-2630/12/5/053002} {\bibfield  {journal}
  {\bibinfo  {journal} {New Journal of Physics}\ }\textbf {\bibinfo {volume}
  {12}},\ \bibinfo {pages} {053002} (\bibinfo {year} {2010})}\BibitemShut
  {NoStop}%
\bibitem [{\citenamefont {Hillery}\ \emph {et~al.}(1999)\citenamefont
  {Hillery}, \citenamefont {Bužek},\ and\ \citenamefont
  {Berthiaume}}]{Hillery1999}%
  \BibitemOpen
  \bibfield  {author} {\bibinfo {author} {\bibfnamefont {M.}~\bibnamefont
  {Hillery}}, \bibinfo {author} {\bibfnamefont {V.}~\bibnamefont {Bužek}},\
  and\ \bibinfo {author} {\bibfnamefont {A.}~\bibnamefont {Berthiaume}},\
  }\bibfield  {title} {\bibinfo {title} {Quantum secret sharing},\ }\href
  {https://doi.org/10.1103/physreva.59.1829} {\bibfield  {journal} {\bibinfo
  {journal} {Physical Review A}\ }\textbf {\bibinfo {volume} {59}},\ \bibinfo
  {pages} {1829} (\bibinfo {year} {1999})}\BibitemShut {NoStop}%
\bibitem [{\citenamefont {Gottesman}(2000)}]{Gottesman2000}%
  \BibitemOpen
  \bibfield  {author} {\bibinfo {author} {\bibfnamefont {D.}~\bibnamefont
  {Gottesman}},\ }\bibfield  {title} {\bibinfo {title} {Theory of quantum
  secret sharing},\ }\href {https://doi.org/10.1103/physreva.61.042311}
  {\bibfield  {journal} {\bibinfo  {journal} {Physical Review A}\ }\textbf
  {\bibinfo {volume} {61}},\ \bibinfo {pages} {042311} (\bibinfo {year}
  {2000})}\BibitemShut {NoStop}%
\bibitem [{\citenamefont {Tittel}\ \emph {et~al.}(2001)\citenamefont {Tittel},
  \citenamefont {Zbinden},\ and\ \citenamefont {Gisin}}]{Tittel2001}%
  \BibitemOpen
  \bibfield  {author} {\bibinfo {author} {\bibfnamefont {W.}~\bibnamefont
  {Tittel}}, \bibinfo {author} {\bibfnamefont {H.}~\bibnamefont {Zbinden}},\
  and\ \bibinfo {author} {\bibfnamefont {N.}~\bibnamefont {Gisin}},\ }\bibfield
   {title} {\bibinfo {title} {Experimental demonstration of quantum secret
  sharing},\ }\href {https://doi.org/10.1103/physreva.63.042301} {\bibfield
  {journal} {\bibinfo  {journal} {Physical Review A}\ }\textbf {\bibinfo
  {volume} {63}},\ \bibinfo {pages} {042301} (\bibinfo {year}
  {2001})}\BibitemShut {NoStop}%
\bibitem [{\citenamefont {Koch}\ \emph {et~al.}(2007)\citenamefont {Koch},
  \citenamefont {Yu}, \citenamefont {Gambetta}, \citenamefont {Houck},
  \citenamefont {Schuster}, \citenamefont {Majer}, \citenamefont {Blais},
  \citenamefont {Devoret}, \citenamefont {Girvin},\ and\ \citenamefont
  {Schoelkopf}}]{Koch2007}%
  \BibitemOpen
  \bibfield  {author} {\bibinfo {author} {\bibfnamefont {J.}~\bibnamefont
  {Koch}}, \bibinfo {author} {\bibfnamefont {T.~M.}\ \bibnamefont {Yu}},
  \bibinfo {author} {\bibfnamefont {J.}~\bibnamefont {Gambetta}}, \bibinfo
  {author} {\bibfnamefont {A.~A.}\ \bibnamefont {Houck}}, \bibinfo {author}
  {\bibfnamefont {D.~I.}\ \bibnamefont {Schuster}}, \bibinfo {author}
  {\bibfnamefont {J.}~\bibnamefont {Majer}}, \bibinfo {author} {\bibfnamefont
  {A.}~\bibnamefont {Blais}}, \bibinfo {author} {\bibfnamefont {M.~H.}\
  \bibnamefont {Devoret}}, \bibinfo {author} {\bibfnamefont {S.~M.}\
  \bibnamefont {Girvin}},\ and\ \bibinfo {author} {\bibfnamefont {R.~J.}\
  \bibnamefont {Schoelkopf}},\ }\bibfield  {title} {\bibinfo {title}
  {Charge-insensitive qubit design derived from the {Cooper} pair box},\ }\href
  {https://doi.org/10.1103/physreva.76.042319} {\bibfield  {journal} {\bibinfo
  {journal} {Physical Review A}\ }\textbf {\bibinfo {volume} {76}},\ \bibinfo
  {pages} {042319} (\bibinfo {year} {2007})}\BibitemShut {NoStop}%
\bibitem [{\citenamefont {Barends}\ \emph {et~al.}(2013)\citenamefont
  {Barends}, \citenamefont {Kelly}, \citenamefont {Megrant}, \citenamefont
  {Sank}, \citenamefont {Jeffrey}, \citenamefont {Chen}, \citenamefont {Yin},
  \citenamefont {Chiaro}, \citenamefont {Mutus}, \citenamefont {Neill},
  \citenamefont {O'Malley}, \citenamefont {Roushan}, \citenamefont {Wenner},
  \citenamefont {White}, \citenamefont {Cleland},\ and\ \citenamefont
  {Martinis}}]{Barends2013}%
  \BibitemOpen
  \bibfield  {author} {\bibinfo {author} {\bibfnamefont {R.}~\bibnamefont
  {Barends}}, \bibinfo {author} {\bibfnamefont {J.}~\bibnamefont {Kelly}},
  \bibinfo {author} {\bibfnamefont {A.}~\bibnamefont {Megrant}}, \bibinfo
  {author} {\bibfnamefont {D.}~\bibnamefont {Sank}}, \bibinfo {author}
  {\bibfnamefont {E.}~\bibnamefont {Jeffrey}}, \bibinfo {author} {\bibfnamefont
  {Y.}~\bibnamefont {Chen}}, \bibinfo {author} {\bibfnamefont {Y.}~\bibnamefont
  {Yin}}, \bibinfo {author} {\bibfnamefont {B.}~\bibnamefont {Chiaro}},
  \bibinfo {author} {\bibfnamefont {J.}~\bibnamefont {Mutus}}, \bibinfo
  {author} {\bibfnamefont {C.}~\bibnamefont {Neill}}, \bibinfo {author}
  {\bibfnamefont {P.}~\bibnamefont {O'Malley}}, \bibinfo {author}
  {\bibfnamefont {P.}~\bibnamefont {Roushan}}, \bibinfo {author} {\bibfnamefont
  {J.}~\bibnamefont {Wenner}}, \bibinfo {author} {\bibfnamefont {T.~C.}\
  \bibnamefont {White}}, \bibinfo {author} {\bibfnamefont {A.~N.}\ \bibnamefont
  {Cleland}},\ and\ \bibinfo {author} {\bibfnamefont {J.~M.}\ \bibnamefont
  {Martinis}},\ }\bibfield  {title} {\bibinfo {title} {Coherent {Josephson}
  qubit suitable for scalable quantum integrated circuits},\ }\href
  {https://doi.org/10.1103/physrevlett.111.080502} {\bibfield  {journal}
  {\bibinfo  {journal} {Physical Review Letters}\ }\textbf {\bibinfo {volume}
  {111}},\ \bibinfo {pages} {080502} (\bibinfo {year} {2013})}\BibitemShut
  {NoStop}%
\bibitem [{\citenamefont {Chen}\ \emph {et~al.}(2014)\citenamefont {Chen},
  \citenamefont {Neill}, \citenamefont {Roushan}, \citenamefont {Leung},
  \citenamefont {Fang}, \citenamefont {Barends}, \citenamefont {Kelly},
  \citenamefont {Campbell}, \citenamefont {Chen}, \citenamefont {Chiaro},
  \citenamefont {Dunsworth}, \citenamefont {Jeffrey}, \citenamefont {Megrant},
  \citenamefont {Mutus}, \citenamefont {O'Malley}, \citenamefont {Quintana},
  \citenamefont {Sank}, \citenamefont {Vainsencher}, \citenamefont {Wenner},
  \citenamefont {White}, \citenamefont {Geller}, \citenamefont {Cleland},\ and\
  \citenamefont {Martinis}}]{Chen2014}%
  \BibitemOpen
  \bibfield  {author} {\bibinfo {author} {\bibfnamefont {Y.}~\bibnamefont
  {Chen}}, \bibinfo {author} {\bibfnamefont {C.}~\bibnamefont {Neill}},
  \bibinfo {author} {\bibfnamefont {P.}~\bibnamefont {Roushan}}, \bibinfo
  {author} {\bibfnamefont {N.}~\bibnamefont {Leung}}, \bibinfo {author}
  {\bibfnamefont {M.}~\bibnamefont {Fang}}, \bibinfo {author} {\bibfnamefont
  {R.}~\bibnamefont {Barends}}, \bibinfo {author} {\bibfnamefont
  {J.}~\bibnamefont {Kelly}}, \bibinfo {author} {\bibfnamefont
  {B.}~\bibnamefont {Campbell}}, \bibinfo {author} {\bibfnamefont
  {Z.}~\bibnamefont {Chen}}, \bibinfo {author} {\bibfnamefont {B.}~\bibnamefont
  {Chiaro}}, \bibinfo {author} {\bibfnamefont {A.}~\bibnamefont {Dunsworth}},
  \bibinfo {author} {\bibfnamefont {E.}~\bibnamefont {Jeffrey}}, \bibinfo
  {author} {\bibfnamefont {A.}~\bibnamefont {Megrant}}, \bibinfo {author}
  {\bibfnamefont {J.}~\bibnamefont {Mutus}}, \bibinfo {author} {\bibfnamefont
  {P.}~\bibnamefont {O'Malley}}, \bibinfo {author} {\bibfnamefont
  {C.}~\bibnamefont {Quintana}}, \bibinfo {author} {\bibfnamefont
  {D.}~\bibnamefont {Sank}}, \bibinfo {author} {\bibfnamefont {A.}~\bibnamefont
  {Vainsencher}}, \bibinfo {author} {\bibfnamefont {J.}~\bibnamefont {Wenner}},
  \bibinfo {author} {\bibfnamefont {T.}~\bibnamefont {White}}, \bibinfo
  {author} {\bibfnamefont {M.~R.}\ \bibnamefont {Geller}}, \bibinfo {author}
  {\bibfnamefont {A.}~\bibnamefont {Cleland}},\ and\ \bibinfo {author}
  {\bibfnamefont {J.~M.}\ \bibnamefont {Martinis}},\ }\bibfield  {title}
  {\bibinfo {title} {Qubit architecture with high coherence and fast tunable
  coupling},\ }\href {https://doi.org/10.1103/physrevlett.113.220502}
  {\bibfield  {journal} {\bibinfo  {journal} {Physical Review Letters}\
  }\textbf {\bibinfo {volume} {113}},\ \bibinfo {pages} {220502} (\bibinfo
  {year} {2014})}\BibitemShut {NoStop}%
\bibitem [{\citenamefont {Jeffrey}\ \emph {et~al.}(2014)\citenamefont
  {Jeffrey}, \citenamefont {Sank}, \citenamefont {Mutus}, \citenamefont
  {White}, \citenamefont {Kelly}, \citenamefont {Barends}, \citenamefont
  {Chen}, \citenamefont {Chen}, \citenamefont {Chiaro}, \citenamefont
  {Dunsworth}, \citenamefont {Megrant}, \citenamefont {O'Malley}, \citenamefont
  {Neill}, \citenamefont {Roushan}, \citenamefont {Vainsencher}, \citenamefont
  {Wenner}, \citenamefont {Cleland},\ and\ \citenamefont
  {Martinis}}]{Jeffrey2014}%
  \BibitemOpen
  \bibfield  {author} {\bibinfo {author} {\bibfnamefont {E.}~\bibnamefont
  {Jeffrey}}, \bibinfo {author} {\bibfnamefont {D.}~\bibnamefont {Sank}},
  \bibinfo {author} {\bibfnamefont {J.~Y.}\ \bibnamefont {Mutus}}, \bibinfo
  {author} {\bibfnamefont {T.~C.}\ \bibnamefont {White}}, \bibinfo {author}
  {\bibfnamefont {J.}~\bibnamefont {Kelly}}, \bibinfo {author} {\bibfnamefont
  {R.}~\bibnamefont {Barends}}, \bibinfo {author} {\bibfnamefont
  {Y.}~\bibnamefont {Chen}}, \bibinfo {author} {\bibfnamefont {Z.}~\bibnamefont
  {Chen}}, \bibinfo {author} {\bibfnamefont {B.}~\bibnamefont {Chiaro}},
  \bibinfo {author} {\bibfnamefont {A.}~\bibnamefont {Dunsworth}}, \bibinfo
  {author} {\bibfnamefont {A.}~\bibnamefont {Megrant}}, \bibinfo {author}
  {\bibfnamefont {P.~J.~J.}\ \bibnamefont {O'Malley}}, \bibinfo {author}
  {\bibfnamefont {C.}~\bibnamefont {Neill}}, \bibinfo {author} {\bibfnamefont
  {P.}~\bibnamefont {Roushan}}, \bibinfo {author} {\bibfnamefont
  {A.}~\bibnamefont {Vainsencher}}, \bibinfo {author} {\bibfnamefont
  {J.}~\bibnamefont {Wenner}}, \bibinfo {author} {\bibfnamefont {A.~N.}\
  \bibnamefont {Cleland}},\ and\ \bibinfo {author} {\bibfnamefont {J.~M.}\
  \bibnamefont {Martinis}},\ }\bibfield  {title} {\bibinfo {title} {Fast
  accurate state measurement with superconducting qubits},\ }\href
  {https://doi.org/10.1103/physrevlett.112.190504} {\bibfield  {journal}
  {\bibinfo  {journal} {Physical Review Letters}\ }\textbf {\bibinfo {volume}
  {112}},\ \bibinfo {pages} {190504} (\bibinfo {year} {2014})}\BibitemShut
  {NoStop}%
\bibitem [{\citenamefont {Yan}\ \emph {et~al.}(2023)\citenamefont {Yan},
  \citenamefont {Wu}, \citenamefont {Lingenfelter}, \citenamefont {Joshi},
  \citenamefont {Andersson}, \citenamefont {Conner}, \citenamefont {Chou},
  \citenamefont {Grebel}, \citenamefont {Miller}, \citenamefont {Povey},
  \citenamefont {Qiao}, \citenamefont {Clerk},\ and\ \citenamefont
  {Cleland}}]{Yan2023}%
  \BibitemOpen
  \bibfield  {author} {\bibinfo {author} {\bibfnamefont {H.}~\bibnamefont
  {Yan}}, \bibinfo {author} {\bibfnamefont {X.}~\bibnamefont {Wu}}, \bibinfo
  {author} {\bibfnamefont {A.}~\bibnamefont {Lingenfelter}}, \bibinfo {author}
  {\bibfnamefont {Y.~J.}\ \bibnamefont {Joshi}}, \bibinfo {author}
  {\bibfnamefont {G.}~\bibnamefont {Andersson}}, \bibinfo {author}
  {\bibfnamefont {C.~R.}\ \bibnamefont {Conner}}, \bibinfo {author}
  {\bibfnamefont {M.-H.}\ \bibnamefont {Chou}}, \bibinfo {author}
  {\bibfnamefont {J.}~\bibnamefont {Grebel}}, \bibinfo {author} {\bibfnamefont
  {J.~M.}\ \bibnamefont {Miller}}, \bibinfo {author} {\bibfnamefont {R.~G.}\
  \bibnamefont {Povey}}, \bibinfo {author} {\bibfnamefont {H.}~\bibnamefont
  {Qiao}}, \bibinfo {author} {\bibfnamefont {A.~A.}\ \bibnamefont {Clerk}},\
  and\ \bibinfo {author} {\bibfnamefont {A.~N.}\ \bibnamefont {Cleland}},\
  }\bibfield  {title} {\bibinfo {title} {Broadband bandpass {Purcell} filter
  for circuit quantum electrodynamics},\ }\href
  {https://doi.org/10.1063/5.0161893} {\bibfield  {journal} {\bibinfo
  {journal} {Applied Physics Letters}\ }\textbf {\bibinfo {volume} {123}},\
  \bibinfo {pages} {134001} (\bibinfo {year} {2023})}\BibitemShut {NoStop}%
\bibitem [{sup()}]{supp}%
  \BibitemOpen
  \href@noop {} {\bibinfo {title} {See {S}upplemental {M}aterial}}\BibitemShut
  {NoStop}%
\bibitem [{\citenamefont {Motzoi}\ \emph {et~al.}(2009)\citenamefont {Motzoi},
  \citenamefont {Gambetta}, \citenamefont {Rebentrost},\ and\ \citenamefont
  {Wilhelm}}]{Motzoi2009}%
  \BibitemOpen
  \bibfield  {author} {\bibinfo {author} {\bibfnamefont {F.}~\bibnamefont
  {Motzoi}}, \bibinfo {author} {\bibfnamefont {J.~M.}\ \bibnamefont
  {Gambetta}}, \bibinfo {author} {\bibfnamefont {P.}~\bibnamefont
  {Rebentrost}},\ and\ \bibinfo {author} {\bibfnamefont {F.~K.}\ \bibnamefont
  {Wilhelm}},\ }\bibfield  {title} {\bibinfo {title} {Simple pulses for
  elimination of leakage in weakly nonlinear qubits},\ }\href
  {https://doi.org/10.1103/physrevlett.103.110501} {\bibfield  {journal}
  {\bibinfo  {journal} {Physical Review Letters}\ }\textbf {\bibinfo {volume}
  {103}},\ \bibinfo {pages} {110501} (\bibinfo {year} {2009})}\BibitemShut
  {NoStop}%
\bibitem [{\citenamefont {Knill}\ \emph {et~al.}(2008)\citenamefont {Knill},
  \citenamefont {Leibfried}, \citenamefont {Reichle}, \citenamefont {Britton},
  \citenamefont {Blakestad}, \citenamefont {Jost}, \citenamefont {Langer},
  \citenamefont {Ozeri}, \citenamefont {Seidelin},\ and\ \citenamefont
  {Wineland}}]{Knill2008}%
  \BibitemOpen
  \bibfield  {author} {\bibinfo {author} {\bibfnamefont {E.}~\bibnamefont
  {Knill}}, \bibinfo {author} {\bibfnamefont {D.}~\bibnamefont {Leibfried}},
  \bibinfo {author} {\bibfnamefont {R.}~\bibnamefont {Reichle}}, \bibinfo
  {author} {\bibfnamefont {J.}~\bibnamefont {Britton}}, \bibinfo {author}
  {\bibfnamefont {R.~B.}\ \bibnamefont {Blakestad}}, \bibinfo {author}
  {\bibfnamefont {J.~D.}\ \bibnamefont {Jost}}, \bibinfo {author}
  {\bibfnamefont {C.}~\bibnamefont {Langer}}, \bibinfo {author} {\bibfnamefont
  {R.}~\bibnamefont {Ozeri}}, \bibinfo {author} {\bibfnamefont
  {S.}~\bibnamefont {Seidelin}},\ and\ \bibinfo {author} {\bibfnamefont
  {D.~J.}\ \bibnamefont {Wineland}},\ }\bibfield  {title} {\bibinfo {title}
  {Randomized benchmarking of quantum gates},\ }\href
  {https://doi.org/10.1103/physreva.77.012307} {\bibfield  {journal} {\bibinfo
  {journal} {Physical Review A}\ }\textbf {\bibinfo {volume} {77}},\ \bibinfo
  {pages} {012307} (\bibinfo {year} {2008})}\BibitemShut {NoStop}%
\bibitem [{\citenamefont {Boixo}\ \emph {et~al.}(2018)\citenamefont {Boixo},
  \citenamefont {Isakov}, \citenamefont {Smelyanskiy}, \citenamefont {Babbush},
  \citenamefont {Ding}, \citenamefont {Jiang}, \citenamefont {Bremner},
  \citenamefont {Martinis},\ and\ \citenamefont {Neven}}]{Boixo2018}%
  \BibitemOpen
  \bibfield  {author} {\bibinfo {author} {\bibfnamefont {S.}~\bibnamefont
  {Boixo}}, \bibinfo {author} {\bibfnamefont {S.~V.}\ \bibnamefont {Isakov}},
  \bibinfo {author} {\bibfnamefont {V.~N.}\ \bibnamefont {Smelyanskiy}},
  \bibinfo {author} {\bibfnamefont {R.}~\bibnamefont {Babbush}}, \bibinfo
  {author} {\bibfnamefont {N.}~\bibnamefont {Ding}}, \bibinfo {author}
  {\bibfnamefont {Z.}~\bibnamefont {Jiang}}, \bibinfo {author} {\bibfnamefont
  {M.~J.}\ \bibnamefont {Bremner}}, \bibinfo {author} {\bibfnamefont {J.~M.}\
  \bibnamefont {Martinis}},\ and\ \bibinfo {author} {\bibfnamefont
  {H.}~\bibnamefont {Neven}},\ }\bibfield  {title} {\bibinfo {title}
  {Characterizing quantum supremacy in near-term devices},\ }\href
  {https://doi.org/10.1038/s41567-018-0124-x} {\bibfield  {journal} {\bibinfo
  {journal} {Nature Physics}\ }\textbf {\bibinfo {volume} {14}},\ \bibinfo
  {pages} {595} (\bibinfo {year} {2018})}\BibitemShut {NoStop}%
\bibitem [{\citenamefont {Shamir}(1979)}]{Shamir1979}%
  \BibitemOpen
  \bibfield  {author} {\bibinfo {author} {\bibfnamefont {A.}~\bibnamefont
  {Shamir}},\ }\bibfield  {title} {\bibinfo {title} {How to share a secret},\
  }\href {https://doi.org/10.1145/359168.359176} {\bibfield  {journal}
  {\bibinfo  {journal} {Communications of the ACM}\ }\textbf {\bibinfo {volume}
  {22}},\ \bibinfo {pages} {612} (\bibinfo {year} {1979})}\BibitemShut
  {NoStop}%
\bibitem [{\citenamefont {Blakley}(1979)}]{Blakley1979}%
  \BibitemOpen
  \bibfield  {author} {\bibinfo {author} {\bibfnamefont {G.~R.}\ \bibnamefont
  {Blakley}},\ }\bibfield  {title} {\bibinfo {title} {Safeguarding
  cryptographic keys},\ }in\ \href {https://doi.org/10.1109/mark.1979.8817296}
  {\emph {\bibinfo {booktitle} {1979 International Workshop on Managing
  Requirements Knowledge (MARK)}}}\ (\bibinfo  {publisher} {IEEE},\ \bibinfo
  {year} {1979})\BibitemShut {NoStop}%
\bibitem [{\citenamefont {Cleve}\ \emph {et~al.}(1999)\citenamefont {Cleve},
  \citenamefont {Gottesman},\ and\ \citenamefont {Lo}}]{Cleve1999}%
  \BibitemOpen
  \bibfield  {author} {\bibinfo {author} {\bibfnamefont {R.}~\bibnamefont
  {Cleve}}, \bibinfo {author} {\bibfnamefont {D.}~\bibnamefont {Gottesman}},\
  and\ \bibinfo {author} {\bibfnamefont {H.-K.}\ \bibnamefont {Lo}},\
  }\bibfield  {title} {\bibinfo {title} {How to share a quantum secret},\
  }\href {https://doi.org/10.1103/physrevlett.83.648} {\bibfield  {journal}
  {\bibinfo  {journal} {Physical Review Letters}\ }\textbf {\bibinfo {volume}
  {83}},\ \bibinfo {pages} {648} (\bibinfo {year} {1999})}\BibitemShut
  {NoStop}%
\bibitem [{\citenamefont {Karlsson}\ \emph {et~al.}(1999)\citenamefont
  {Karlsson}, \citenamefont {Koashi},\ and\ \citenamefont
  {Imoto}}]{Karlsson1999}%
  \BibitemOpen
  \bibfield  {author} {\bibinfo {author} {\bibfnamefont {A.}~\bibnamefont
  {Karlsson}}, \bibinfo {author} {\bibfnamefont {M.}~\bibnamefont {Koashi}},\
  and\ \bibinfo {author} {\bibfnamefont {N.}~\bibnamefont {Imoto}},\ }\bibfield
   {title} {\bibinfo {title} {Quantum entanglement for secret sharing and
  secret splitting},\ }\href {https://doi.org/10.1103/physreva.59.162}
  {\bibfield  {journal} {\bibinfo  {journal} {Physical Review A}\ }\textbf
  {\bibinfo {volume} {59}},\ \bibinfo {pages} {162} (\bibinfo {year}
  {1999})}\BibitemShut {NoStop}%
\bibitem [{\citenamefont {Guo}\ and\ \citenamefont {Guo}(2003)}]{Guo2003}%
  \BibitemOpen
  \bibfield  {author} {\bibinfo {author} {\bibfnamefont {G.-P.}\ \bibnamefont
  {Guo}}\ and\ \bibinfo {author} {\bibfnamefont {G.-C.}\ \bibnamefont {Guo}},\
  }\bibfield  {title} {\bibinfo {title} {Quantum secret sharing without
  entanglement},\ }\href {https://doi.org/10.1016/s0375-9601(03)00074-4}
  {\bibfield  {journal} {\bibinfo  {journal} {Physics Letters A}\ }\textbf
  {\bibinfo {volume} {310}},\ \bibinfo {pages} {247} (\bibinfo {year}
  {2003})}\BibitemShut {NoStop}%
\bibitem [{\citenamefont {Xiao}\ \emph {et~al.}(2004)\citenamefont {Xiao},
  \citenamefont {Lu~Long}, \citenamefont {Deng},\ and\ \citenamefont
  {Pan}}]{Xiao2004}%
  \BibitemOpen
  \bibfield  {author} {\bibinfo {author} {\bibfnamefont {L.}~\bibnamefont
  {Xiao}}, \bibinfo {author} {\bibfnamefont {G.}~\bibnamefont {Lu~Long}},
  \bibinfo {author} {\bibfnamefont {F.-G.}\ \bibnamefont {Deng}},\ and\
  \bibinfo {author} {\bibfnamefont {J.-W.}\ \bibnamefont {Pan}},\ }\bibfield
  {title} {\bibinfo {title} {Efficient multiparty quantum-secret-sharing
  schemes},\ }\href {https://doi.org/10.1103/physreva.69.052307} {\bibfield
  {journal} {\bibinfo  {journal} {Physical Review A}\ }\textbf {\bibinfo
  {volume} {69}},\ \bibinfo {pages} {052307} (\bibinfo {year}
  {2004})}\BibitemShut {NoStop}%
\bibitem [{\citenamefont {Zhang}\ \emph {et~al.}(2005)\citenamefont {Zhang},
  \citenamefont {Li},\ and\ \citenamefont {Man}}]{Zhang2005}%
  \BibitemOpen
  \bibfield  {author} {\bibinfo {author} {\bibfnamefont {Z.-j.}\ \bibnamefont
  {Zhang}}, \bibinfo {author} {\bibfnamefont {Y.}~\bibnamefont {Li}},\ and\
  \bibinfo {author} {\bibfnamefont {Z.-x.}\ \bibnamefont {Man}},\ }\bibfield
  {title} {\bibinfo {title} {Multiparty quantum secret sharing},\ }\href
  {https://doi.org/10.1103/physreva.71.044301} {\bibfield  {journal} {\bibinfo
  {journal} {Physical Review A}\ }\textbf {\bibinfo {volume} {71}},\ \bibinfo
  {pages} {044301} (\bibinfo {year} {2005})}\BibitemShut {NoStop}%
\bibitem [{\citenamefont {Zhang}\ and\ \citenamefont {Man}(2005)}]{Zhang2005a}%
  \BibitemOpen
  \bibfield  {author} {\bibinfo {author} {\bibfnamefont {Z.-j.}\ \bibnamefont
  {Zhang}}\ and\ \bibinfo {author} {\bibfnamefont {Z.-x.}\ \bibnamefont
  {Man}},\ }\bibfield  {title} {\bibinfo {title} {Multiparty quantum secret
  sharing of classical messages based on entanglement swapping},\ }\href
  {https://doi.org/10.1103/physreva.72.022303} {\bibfield  {journal} {\bibinfo
  {journal} {Physical Review A}\ }\textbf {\bibinfo {volume} {72}},\ \bibinfo
  {pages} {022303} (\bibinfo {year} {2005})}\BibitemShut {NoStop}%
\bibitem [{\citenamefont {Markham}\ and\ \citenamefont
  {Sanders}(2008)}]{Markham2008}%
  \BibitemOpen
  \bibfield  {author} {\bibinfo {author} {\bibfnamefont {D.}~\bibnamefont
  {Markham}}\ and\ \bibinfo {author} {\bibfnamefont {B.~C.}\ \bibnamefont
  {Sanders}},\ }\bibfield  {title} {\bibinfo {title} {Graph states for quantum
  secret sharing},\ }\href {https://doi.org/10.1103/physreva.78.042309}
  {\bibfield  {journal} {\bibinfo  {journal} {Physical Review A}\ }\textbf
  {\bibinfo {volume} {78}},\ \bibinfo {pages} {042309} (\bibinfo {year}
  {2008})}\BibitemShut {NoStop}%
\bibitem [{\citenamefont {Chen}\ \emph {et~al.}(2005)\citenamefont {Chen},
  \citenamefont {Zhang}, \citenamefont {Zhao}, \citenamefont {Zhou},
  \citenamefont {Lu}, \citenamefont {Peng}, \citenamefont {Yang},\ and\
  \citenamefont {Pan}}]{Chen2005}%
  \BibitemOpen
  \bibfield  {author} {\bibinfo {author} {\bibfnamefont {Y.-A.}\ \bibnamefont
  {Chen}}, \bibinfo {author} {\bibfnamefont {A.-N.}\ \bibnamefont {Zhang}},
  \bibinfo {author} {\bibfnamefont {Z.}~\bibnamefont {Zhao}}, \bibinfo {author}
  {\bibfnamefont {X.-Q.}\ \bibnamefont {Zhou}}, \bibinfo {author}
  {\bibfnamefont {C.-Y.}\ \bibnamefont {Lu}}, \bibinfo {author} {\bibfnamefont
  {C.-Z.}\ \bibnamefont {Peng}}, \bibinfo {author} {\bibfnamefont
  {T.}~\bibnamefont {Yang}},\ and\ \bibinfo {author} {\bibfnamefont {J.-W.}\
  \bibnamefont {Pan}},\ }\bibfield  {title} {\bibinfo {title} {Experimental
  quantum secret sharing and third-man quantum cryptography},\ }\href
  {https://doi.org/10.1103/physrevlett.95.200502} {\bibfield  {journal}
  {\bibinfo  {journal} {Physical Review Letters}\ }\textbf {\bibinfo {volume}
  {95}},\ \bibinfo {pages} {200502} (\bibinfo {year} {2005})}\BibitemShut
  {NoStop}%
\bibitem [{\citenamefont {Gaertner}\ \emph {et~al.}(2007)\citenamefont
  {Gaertner}, \citenamefont {Kurtsiefer}, \citenamefont {Bourennane},\ and\
  \citenamefont {Weinfurter}}]{Gaertner2007}%
  \BibitemOpen
  \bibfield  {author} {\bibinfo {author} {\bibfnamefont {S.}~\bibnamefont
  {Gaertner}}, \bibinfo {author} {\bibfnamefont {C.}~\bibnamefont
  {Kurtsiefer}}, \bibinfo {author} {\bibfnamefont {M.}~\bibnamefont
  {Bourennane}},\ and\ \bibinfo {author} {\bibfnamefont {H.}~\bibnamefont
  {Weinfurter}},\ }\bibfield  {title} {\bibinfo {title} {Experimental
  demonstration of four-party quantum secret sharing},\ }\href
  {https://doi.org/10.1103/physrevlett.98.020503} {\bibfield  {journal}
  {\bibinfo  {journal} {Physical Review Letters}\ }\textbf {\bibinfo {volume}
  {98}},\ \bibinfo {pages} {020503} (\bibinfo {year} {2007})}\BibitemShut
  {NoStop}%
\bibitem [{\citenamefont {Bogdanski}\ \emph {et~al.}(2008)\citenamefont
  {Bogdanski}, \citenamefont {Rafiei},\ and\ \citenamefont
  {Bourennane}}]{Bogdanski2008}%
  \BibitemOpen
  \bibfield  {author} {\bibinfo {author} {\bibfnamefont {J.}~\bibnamefont
  {Bogdanski}}, \bibinfo {author} {\bibfnamefont {N.}~\bibnamefont {Rafiei}},\
  and\ \bibinfo {author} {\bibfnamefont {M.}~\bibnamefont {Bourennane}},\
  }\bibfield  {title} {\bibinfo {title} {Experimental quantum secret sharing
  using telecommunication fiber},\ }\href
  {https://doi.org/10.1103/physreva.78.062307} {\bibfield  {journal} {\bibinfo
  {journal} {Physical Review A}\ }\textbf {\bibinfo {volume} {78}},\ \bibinfo
  {pages} {062307} (\bibinfo {year} {2008})}\BibitemShut {NoStop}%
\bibitem [{\citenamefont {Bell}\ \emph {et~al.}(2014)\citenamefont {Bell},
  \citenamefont {Markham}, \citenamefont {Herrera-Martí}, \citenamefont
  {Marin}, \citenamefont {Wadsworth}, \citenamefont {Rarity},\ and\
  \citenamefont {Tame}}]{Bell2014}%
  \BibitemOpen
  \bibfield  {author} {\bibinfo {author} {\bibfnamefont {B.~A.}\ \bibnamefont
  {Bell}}, \bibinfo {author} {\bibfnamefont {D.}~\bibnamefont {Markham}},
  \bibinfo {author} {\bibfnamefont {D.~A.}\ \bibnamefont {Herrera-Martí}},
  \bibinfo {author} {\bibfnamefont {A.}~\bibnamefont {Marin}}, \bibinfo
  {author} {\bibfnamefont {W.~J.}\ \bibnamefont {Wadsworth}}, \bibinfo {author}
  {\bibfnamefont {J.~G.}\ \bibnamefont {Rarity}},\ and\ \bibinfo {author}
  {\bibfnamefont {M.~S.}\ \bibnamefont {Tame}},\ }\bibfield  {title} {\bibinfo
  {title} {Experimental demonstration of graph-state quantum secret sharing},\
  }\href {https://doi.org/10.1038/ncomms6480} {\bibfield  {journal} {\bibinfo
  {journal} {Nature Communications}\ }\textbf {\bibinfo {volume} {5}},\
  \bibinfo {pages} {5480} (\bibinfo {year} {2014})}\BibitemShut {NoStop}%
\bibitem [{\citenamefont {Xiao}\ \emph {et~al.}(2024)\citenamefont {Xiao},
  \citenamefont {Jia}, \citenamefont {Song}, \citenamefont {Bao}, \citenamefont
  {Fu}, \citenamefont {Yin},\ and\ \citenamefont {Chen}}]{Xiao2024}%
  \BibitemOpen
  \bibfield  {author} {\bibinfo {author} {\bibfnamefont {Y.-R.}\ \bibnamefont
  {Xiao}}, \bibinfo {author} {\bibfnamefont {Z.-Y.}\ \bibnamefont {Jia}},
  \bibinfo {author} {\bibfnamefont {Y.-C.}\ \bibnamefont {Song}}, \bibinfo
  {author} {\bibfnamefont {Y.}~\bibnamefont {Bao}}, \bibinfo {author}
  {\bibfnamefont {Y.}~\bibnamefont {Fu}}, \bibinfo {author} {\bibfnamefont
  {H.-L.}\ \bibnamefont {Yin}},\ and\ \bibinfo {author} {\bibfnamefont {Z.-B.}\
  \bibnamefont {Chen}},\ }\bibfield  {title} {\bibinfo {title}
  {Source-independent quantum secret sharing with entangled photon pair
  networks},\ }\href {https://doi.org/10.1364/ol.527857} {\bibfield  {journal}
  {\bibinfo  {journal} {Optics Letters}\ }\textbf {\bibinfo {volume} {49}},\
  \bibinfo {pages} {4210} (\bibinfo {year} {2024})}\BibitemShut {NoStop}%
\bibitem [{\citenamefont {Zhang}\ \emph {et~al.}(2024)\citenamefont {Zhang},
  \citenamefont {Ying}, \citenamefont {Wang}, \citenamefont {Zhong},
  \citenamefont {Du}, \citenamefont {Shen}, \citenamefont {Li}, \citenamefont
  {Zhang}, \citenamefont {Gu}, \citenamefont {Wang}, \citenamefont {Zhou},\
  and\ \citenamefont {Sheng}}]{Zhang2024}%
  \BibitemOpen
  \bibfield  {author} {\bibinfo {author} {\bibfnamefont {Q.}~\bibnamefont
  {Zhang}}, \bibinfo {author} {\bibfnamefont {J.-W.}\ \bibnamefont {Ying}},
  \bibinfo {author} {\bibfnamefont {Z.-J.}\ \bibnamefont {Wang}}, \bibinfo
  {author} {\bibfnamefont {W.}~\bibnamefont {Zhong}}, \bibinfo {author}
  {\bibfnamefont {M.-M.}\ \bibnamefont {Du}}, \bibinfo {author} {\bibfnamefont
  {S.-T.}\ \bibnamefont {Shen}}, \bibinfo {author} {\bibfnamefont {X.-Y.}\
  \bibnamefont {Li}}, \bibinfo {author} {\bibfnamefont {A.-L.}\ \bibnamefont
  {Zhang}}, \bibinfo {author} {\bibfnamefont {S.-P.}\ \bibnamefont {Gu}},
  \bibinfo {author} {\bibfnamefont {X.-F.}\ \bibnamefont {Wang}}, \bibinfo
  {author} {\bibfnamefont {L.}~\bibnamefont {Zhou}},\ and\ \bibinfo {author}
  {\bibfnamefont {Y.-B.}\ \bibnamefont {Sheng}},\ }\bibfield  {title} {\bibinfo
  {title} {Device-independent quantum secret sharing with advanced random key
  generation basis},\ }\bibfield  {journal} {\bibinfo  {journal} {arXiv
  preprint}\ }\href {https://doi.org/10.48550/arXiv.2410.04003}
  {10.48550/arXiv.2410.04003} (\bibinfo {year} {2024})\BibitemShut {NoStop}%
\bibitem [{\citenamefont {Wang}\ \emph {et~al.}(2024)\citenamefont {Wang},
  \citenamefont {Sun}, \citenamefont {Cao}, \citenamefont {Yin},\ and\
  \citenamefont {Chen}}]{Wang2024}%
  \BibitemOpen
  \bibfield  {author} {\bibinfo {author} {\bibfnamefont {Y.-Z.}\ \bibnamefont
  {Wang}}, \bibinfo {author} {\bibfnamefont {X.-R.}\ \bibnamefont {Sun}},
  \bibinfo {author} {\bibfnamefont {X.-Y.}\ \bibnamefont {Cao}}, \bibinfo
  {author} {\bibfnamefont {H.-L.}\ \bibnamefont {Yin}},\ and\ \bibinfo {author}
  {\bibfnamefont {Z.-B.}\ \bibnamefont {Chen}},\ }\bibfield  {title} {\bibinfo
  {title} {Experimental coherent-state quantum secret sharing with finite
  pulses},\ }\href {https://doi.org/10.1103/physrevapplied.22.044018}
  {\bibfield  {journal} {\bibinfo  {journal} {Physical Review Applied}\
  }\textbf {\bibinfo {volume} {22}},\ \bibinfo {pages} {044018} (\bibinfo
  {year} {2024})}\BibitemShut {NoStop}%
\bibitem [{\citenamefont {Zhang}\ \emph {et~al.}(2025)\citenamefont {Zhang},
  \citenamefont {Zhang}, \citenamefont {Zhong}, \citenamefont {Du},
  \citenamefont {Zhou},\ and\ \citenamefont {Sheng}}]{Zhang2025}%
  \BibitemOpen
  \bibfield  {author} {\bibinfo {author} {\bibfnamefont {Q.}~\bibnamefont
  {Zhang}}, \bibinfo {author} {\bibfnamefont {C.}~\bibnamefont {Zhang}},
  \bibinfo {author} {\bibfnamefont {W.}~\bibnamefont {Zhong}}, \bibinfo
  {author} {\bibfnamefont {M.-M.}\ \bibnamefont {Du}}, \bibinfo {author}
  {\bibfnamefont {L.}~\bibnamefont {Zhou}},\ and\ \bibinfo {author}
  {\bibfnamefont {Y.-B.}\ \bibnamefont {Sheng}},\ }\bibfield  {title} {\bibinfo
  {title} {High-efficient long-distance device-independent quantum secret
  sharing based on single-photon sources},\ }\bibfield  {journal} {\bibinfo
  {journal} {arXiv preprint}\ }\href
  {https://doi.org/10.48550/arXiv.2505.10797} {10.48550/arXiv.2505.10797}
  (\bibinfo {year} {2025})\BibitemShut {NoStop}%
\bibitem [{\citenamefont {Basak}\ \emph {et~al.}(2023)\citenamefont {Basak},
  \citenamefont {Das}, \citenamefont {Paul}, \citenamefont {Nandi},\ and\
  \citenamefont {Patel}}]{Basak2023}%
  \BibitemOpen
  \bibfield  {author} {\bibinfo {author} {\bibfnamefont {N.}~\bibnamefont
  {Basak}}, \bibinfo {author} {\bibfnamefont {N.}~\bibnamefont {Das}}, \bibinfo
  {author} {\bibfnamefont {G.}~\bibnamefont {Paul}}, \bibinfo {author}
  {\bibfnamefont {K.}~\bibnamefont {Nandi}},\ and\ \bibinfo {author}
  {\bibfnamefont {N.}~\bibnamefont {Patel}},\ }\bibfield  {title} {\bibinfo
  {title} {Quantum secret sharing protocol using {GHZ} state: implementation on
  {IBM} qiskit},\ }\href {https://doi.org/10.1007/s11128-023-04129-4}
  {\bibfield  {journal} {\bibinfo  {journal} {Quantum Information Processing}\
  }\textbf {\bibinfo {volume} {22}},\ \bibinfo {pages} {393} (\bibinfo {year}
  {2023})}\BibitemShut {NoStop}%
\bibitem [{\citenamefont {Graves}\ \emph {et~al.}(2024)\citenamefont {Graves},
  \citenamefont {Nelson},\ and\ \citenamefont {Chitambar}}]{graves2024}%
  \BibitemOpen
  \bibfield  {author} {\bibinfo {author} {\bibfnamefont {J.}~\bibnamefont
  {Graves}}, \bibinfo {author} {\bibfnamefont {M.}~\bibnamefont {Nelson}},\
  and\ \bibinfo {author} {\bibfnamefont {E.}~\bibnamefont {Chitambar}},\
  }\bibfield  {title} {\bibinfo {title} {Implementing quantum secret sharing on
  current hardware},\ }\bibfield  {journal} {\bibinfo  {journal} {arXiv
  preprint}\ }\href {https://doi.org/10.48550/arXiv.2410.11640}
  {10.48550/arXiv.2410.11640} (\bibinfo {year} {2024})\BibitemShut {NoStop}%
\bibitem [{\citenamefont {Greenberger}\ \emph {et~al.}(1989)\citenamefont
  {Greenberger}, \citenamefont {Horne},\ and\ \citenamefont
  {Zeilinger}}]{Greenberger1989}%
  \BibitemOpen
  \bibfield  {author} {\bibinfo {author} {\bibfnamefont {D.~M.}\ \bibnamefont
  {Greenberger}}, \bibinfo {author} {\bibfnamefont {M.~A.}\ \bibnamefont
  {Horne}},\ and\ \bibinfo {author} {\bibfnamefont {A.}~\bibnamefont
  {Zeilinger}},\ }\bibinfo {title} {Going beyond {B}ell's theorem},\ in\
  \href {https://doi.org/10.1007/978-94-017-0849-4_10} {\emph {\bibinfo
  {booktitle} {Bell's Theorem, Quantum Theory and Conceptions of the
  Universe}}}\ (\bibinfo  {publisher} {Springer Netherlands},\ \bibinfo {year}
  {1989})\ pp.\ \bibinfo {pages} {69--72}\BibitemShut {NoStop}%
\bibitem [{\citenamefont {Barnett}\ and\ \citenamefont
  {Croke}(2009)}]{Barnett2009}%
  \BibitemOpen
  \bibfield  {author} {\bibinfo {author} {\bibfnamefont {S.~M.}\ \bibnamefont
  {Barnett}}\ and\ \bibinfo {author} {\bibfnamefont {S.}~\bibnamefont
  {Croke}},\ }\bibfield  {title} {\bibinfo {title} {Quantum state
  discrimination},\ }\href {https://doi.org/10.1364/aop.1.000238} {\bibfield
  {journal} {\bibinfo  {journal} {Advances in Optics and Photonics}\ }\textbf
  {\bibinfo {volume} {1}},\ \bibinfo {pages} {238} (\bibinfo {year}
  {2009})}\BibitemShut {NoStop}%
\bibitem [{\citenamefont {Park}(1970)}]{Park1970}%
  \BibitemOpen
  \bibfield  {author} {\bibinfo {author} {\bibfnamefont {J.~L.}\ \bibnamefont
  {Park}},\ }\bibfield  {title} {\bibinfo {title} {The concept of transition in
  quantum mechanics},\ }\href {https://doi.org/10.1007/bf00708652} {\bibfield
  {journal} {\bibinfo  {journal} {Foundations of Physics}\ }\textbf {\bibinfo
  {volume} {1}},\ \bibinfo {pages} {23} (\bibinfo {year} {1970})}\BibitemShut
  {NoStop}%
\bibitem [{\citenamefont {Devetak}\ and\ \citenamefont
  {Winter}(2004)}]{Devetak2004}%
  \BibitemOpen
  \bibfield  {author} {\bibinfo {author} {\bibfnamefont {I.}~\bibnamefont
  {Devetak}}\ and\ \bibinfo {author} {\bibfnamefont {A.}~\bibnamefont
  {Winter}},\ }\bibfield  {title} {\bibinfo {title} {Relating quantum privacy
  and quantum coherence: an operational approach},\ }\href
  {https://doi.org/10.1103/PhysRevLett.93.080501} {\bibfield  {journal}
  {\bibinfo  {journal} {Physical Review Letters}\ }\textbf {\bibinfo {volume}
  {93}},\ \bibinfo {pages} {080501} (\bibinfo {year} {2004})}\BibitemShut
  {NoStop}%
\bibitem [{\citenamefont {Dür}\ \emph {et~al.}(2003)\citenamefont {Dür},
  \citenamefont {Aschauer},\ and\ \citenamefont {Briegel}}]{Duer2003}%
  \BibitemOpen
  \bibfield  {author} {\bibinfo {author} {\bibfnamefont {W.}~\bibnamefont
  {Dür}}, \bibinfo {author} {\bibfnamefont {H.}~\bibnamefont {Aschauer}},\
  and\ \bibinfo {author} {\bibfnamefont {H.-J.}\ \bibnamefont {Briegel}},\
  }\bibfield  {title} {\bibinfo {title} {Multiparticle entanglement
  purification for graph states},\ }\href
  {https://doi.org/10.1103/physrevlett.91.107903} {\bibfield  {journal}
  {\bibinfo  {journal} {Physical Review Letters}\ }\textbf {\bibinfo {volume}
  {91}},\ \bibinfo {pages} {107903} (\bibinfo {year} {2003})}\BibitemShut
  {NoStop}%
\bibitem [{\citenamefont {Murao}\ \emph {et~al.}(1998)\citenamefont {Murao},
  \citenamefont {Plenio}, \citenamefont {Popescu}, \citenamefont {Vedral},\
  and\ \citenamefont {Knight}}]{Murao1998}%
  \BibitemOpen
  \bibfield  {author} {\bibinfo {author} {\bibfnamefont {M.}~\bibnamefont
  {Murao}}, \bibinfo {author} {\bibfnamefont {M.~B.}\ \bibnamefont {Plenio}},
  \bibinfo {author} {\bibfnamefont {S.}~\bibnamefont {Popescu}}, \bibinfo
  {author} {\bibfnamefont {V.}~\bibnamefont {Vedral}},\ and\ \bibinfo {author}
  {\bibfnamefont {P.~L.}\ \bibnamefont {Knight}},\ }\bibfield  {title}
  {\bibinfo {title} {Multiparticle entanglement purification protocols},\
  }\href {https://doi.org/10.1103/physreva.57.r4075} {\bibfield  {journal}
  {\bibinfo  {journal} {Physical Review A}\ }\textbf {\bibinfo {volume} {57}},\
  \bibinfo {pages} {R4075} (\bibinfo {year} {1998})}\BibitemShut {NoStop}%
\bibitem [{\citenamefont {Hillery}\ \emph {et~al.}(2006)\citenamefont
  {Hillery}, \citenamefont {Ziman}, \citenamefont {Bužek},\ and\ \citenamefont
  {Bieliková}}]{Hillery2006}%
  \BibitemOpen
  \bibfield  {author} {\bibinfo {author} {\bibfnamefont {M.}~\bibnamefont
  {Hillery}}, \bibinfo {author} {\bibfnamefont {M.}~\bibnamefont {Ziman}},
  \bibinfo {author} {\bibfnamefont {V.}~\bibnamefont {Bužek}},\ and\ \bibinfo
  {author} {\bibfnamefont {M.}~\bibnamefont {Bieliková}},\ }\bibfield  {title}
  {\bibinfo {title} {Towards quantum-based privacy and voting},\ }\href
  {https://doi.org/10.1016/j.physleta.2005.09.010} {\bibfield  {journal}
  {\bibinfo  {journal} {Physics Letters A}\ }\textbf {\bibinfo {volume}
  {349}},\ \bibinfo {pages} {75} (\bibinfo {year} {2006})}\BibitemShut
  {NoStop}%
\bibitem [{\citenamefont {Xue}\ and\ \citenamefont {Zhang}(2017)}]{Xue2017}%
  \BibitemOpen
  \bibfield  {author} {\bibinfo {author} {\bibfnamefont {P.}~\bibnamefont
  {Xue}}\ and\ \bibinfo {author} {\bibfnamefont {X.}~\bibnamefont {Zhang}},\
  }\bibfield  {title} {\bibinfo {title} {A simple quantum voting scheme with
  multi-qubit entanglement},\ }\href
  {https://doi.org/10.1038/s41598-017-07976-1} {\bibfield  {journal} {\bibinfo
  {journal} {Scientific Reports}\ }\textbf {\bibinfo {volume} {7}},\ \bibinfo
  {pages} {7586} (\bibinfo {year} {2017})}\BibitemShut {NoStop}%
\bibitem [{\citenamefont {Lamport}\ \emph {et~al.}(1982)\citenamefont
  {Lamport}, \citenamefont {Shostak},\ and\ \citenamefont
  {Pease}}]{Lamport1982}%
  \BibitemOpen
  \bibfield  {author} {\bibinfo {author} {\bibfnamefont {L.}~\bibnamefont
  {Lamport}}, \bibinfo {author} {\bibfnamefont {R.}~\bibnamefont {Shostak}},\
  and\ \bibinfo {author} {\bibfnamefont {M.}~\bibnamefont {Pease}},\ }\bibfield
   {title} {\bibinfo {title} {The {B}yzantine generals problem},\ }\href
  {https://doi.org/10.1145/357172.357176} {\bibfield  {journal} {\bibinfo
  {journal} {ACM Transactions on Programming Languages and Systems}\ }\textbf
  {\bibinfo {volume} {4}},\ \bibinfo {pages} {382} (\bibinfo {year}
  {1982})}\BibitemShut {NoStop}%
\bibitem [{\citenamefont {Renou}\ \emph {et~al.}(2019)\citenamefont {Renou},
  \citenamefont {Bäumer}, \citenamefont {Boreiri}, \citenamefont {Brunner},
  \citenamefont {Gisin},\ and\ \citenamefont {Beigi}}]{Renou2019}%
  \BibitemOpen
  \bibfield  {author} {\bibinfo {author} {\bibfnamefont {M.-O.}\ \bibnamefont
  {Renou}}, \bibinfo {author} {\bibfnamefont {E.}~\bibnamefont {Bäumer}},
  \bibinfo {author} {\bibfnamefont {S.}~\bibnamefont {Boreiri}}, \bibinfo
  {author} {\bibfnamefont {N.}~\bibnamefont {Brunner}}, \bibinfo {author}
  {\bibfnamefont {N.}~\bibnamefont {Gisin}},\ and\ \bibinfo {author}
  {\bibfnamefont {S.}~\bibnamefont {Beigi}},\ }\bibfield  {title} {\bibinfo
  {title} {Genuine quantum nonlocality in the triangle network},\ }\href
  {https://doi.org/10.1103/physrevlett.123.140401} {\bibfield  {journal}
  {\bibinfo  {journal} {Physical Review Letters}\ }\textbf {\bibinfo {volume}
  {123}},\ \bibinfo {pages} {140401} (\bibinfo {year} {2019})}\BibitemShut
  {NoStop}%
\bibitem [{\citenamefont {Tavakoli}\ \emph {et~al.}(2022)\citenamefont
  {Tavakoli}, \citenamefont {Pozas-Kerstjens}, \citenamefont {Luo},\ and\
  \citenamefont {Renou}}]{Tavakoli2022}%
  \BibitemOpen
  \bibfield  {author} {\bibinfo {author} {\bibfnamefont {A.}~\bibnamefont
  {Tavakoli}}, \bibinfo {author} {\bibfnamefont {A.}~\bibnamefont
  {Pozas-Kerstjens}}, \bibinfo {author} {\bibfnamefont {M.-X.}\ \bibnamefont
  {Luo}},\ and\ \bibinfo {author} {\bibfnamefont {M.-O.}\ \bibnamefont
  {Renou}},\ }\bibfield  {title} {\bibinfo {title} {Bell nonlocality in
  networks},\ }\href {https://doi.org/10.1088/1361-6633/ac41bb} {\bibfield
  {journal} {\bibinfo  {journal} {Reports on Progress in Physics}\ }\textbf
  {\bibinfo {volume} {85}},\ \bibinfo {pages} {056001} (\bibinfo {year}
  {2022})}\BibitemShut {NoStop}%
\end{thebibliography}%


\begin{thebibliography}{10}

\bibitem{Grebel2024}
Joel Grebel, Haoxiong Yan, Ming-Han Chou, Gustav Andersson, Christopher~R.
  Conner, Yash~J. Joshi, Jacob~M. Miller, Rhys~G. Povey, Hong Qiao, Xuntao Wu,
  and Andrew~N. Cleland.
\newblock Bidirectional multiphoton communication between remote
  superconducting nodes.
\newblock {\em Physical Review Letters}, 132(4):047001, January 2024.

\bibitem{Dolan1977}
G.~J. Dolan.
\newblock Offset masks for lift-off photoprocessing.
\newblock {\em Applied Physics Letters}, 31(5):337--339, September 1977.

\bibitem{Yan2024}
Haoxiong Yan.
\newblock {\em Quantum networking with superconducting qubits}.
\newblock PhD thesis, The University of Chicago, August 2024.

\bibitem{Motzoi2009}
F.~Motzoi, J.~M. Gambetta, P.~Rebentrost, and F.~K. Wilhelm.
\newblock Simple pulses for elimination of leakage in weakly nonlinear qubits.
\newblock {\em Physical Review Letters}, 103(11):110501, September 2009.

\bibitem{Knill2008}
E.~Knill, D.~Leibfried, R.~Reichle, J.~Britton, R.~B. Blakestad, J.~D. Jost,
  C.~Langer, R.~Ozeri, S.~Seidelin, and D.~J. Wineland.
\newblock Randomized benchmarking of quantum gates.
\newblock {\em Physical Review A}, 77(1):012307, January 2008.

\bibitem{Boixo2018}
Sergio Boixo, Sergei~V. Isakov, Vadim~N. Smelyanskiy, Ryan Babbush, Nan Ding,
  Zhang Jiang, Michael~J. Bremner, John~M. Martinis, and Hartmut Neven.
\newblock Characterizing quantum supremacy in near-term devices.
\newblock {\em Nature Physics}, 14(6):595--600, April 2018.

\bibitem{Zhong2019}
Y.~P. Zhong, H.-S. Chang, K.~J. Satzinger, M.-H. Chou, A.~Bienfait, C.~R.
  Conner, É. Dumur, J.~Grebel, G.~A. Peairs, R.~G. Povey, D.~I. Schuster, and
  A.~N. Cleland.
\newblock Violating {Bell’s} inequality with remotely connected
  superconducting qubits.
\newblock {\em Nature Physics}, 15(8):741--744, April 2019.

\bibitem{Bialczak2010}
R.~C. Bialczak, M.~Ansmann, M.~Hofheinz, E.~Lucero, M.~Neeley, A.~D.
  O’Connell, D.~Sank, H.~Wang, J.~Wenner, M.~Steffen, A.~N. Cleland, and
  J.~M. Martinis.
\newblock Quantum process tomography of a universal entangling gate implemented
  with josephson phase qubits.
\newblock {\em Nature Physics}, 6(6):409--413, April 2010.

\bibitem{Zukowski1993}
M.~{\.Z}ukowski, A.~Zeilinger, M.~A. Horne, and A.~K. Ekert.
\newblock ``{E}vent-ready-detectors'' {B}ell experiment via entanglement
  swapping.
\newblock {\em Physical Review Letters}, 71(26):4287--4290, December 1993.

\bibitem{Schumacher1998}
Benjamin Schumacher and Michael~D. Westmoreland.
\newblock Quantum privacy and quantum coherence.
\newblock {\em Physical Review Letters}, 80:5695--5697, June 1998.

\bibitem{Holevo1998}
A.~S. Holevo.
\newblock The capacity of the quantum channel with general signal states.
\newblock {\em IEEE Transactions on Information Theory}, 44(1):269--273, 1998.

\bibitem{Schumacher1997}
Benjamin Schumacher and Michael~D. Westmoreland.
\newblock Sending classical information via noisy quantum channels.
\newblock {\em Physical Review A}, 56(1):131--138, July 1997.

\bibitem{Devetak2004}
I.~Devetak and A.~Winter.
\newblock Relating quantum privacy and quantum coherence: an operational
  approach.
\newblock {\em Physical Review Letters}, 93:080501, August 2004.

\bibitem{Devetak2005}
Igor Devetak and Andreas Winter.
\newblock Distillation of secret key and entanglement from quantum states.
\newblock {\em Proceedings of the Royal Society A: Mathematical, Physical and
  Engineering Sciences}, 461(2053):207--235, January 2005.

\end{thebibliography}
\end{document}


\title{Supplementary Information for ``Quantum secret sharing in a triangular superconducting quantum network''}

\author{Haoxiong Yan}
\altaffiliation[Present address: ]{Applied Materials, Inc, Santa Clara, CA 95051, USA}
\affiliation{Pritzker School of Molecular Engineering, University of Chicago, Chicago IL 60637, USA}

\author{Allen Zang}
\affiliation{Pritzker School of Molecular Engineering, University of Chicago, Chicago IL 60637, USA}

\author{Joel Grebel}
\altaffiliation[Present address: ]{Google Quantum AI, 301 Mentor Dr, Goleta, CA 93111, USA}
\affiliation{Pritzker School of Molecular Engineering, University of Chicago, Chicago IL 60637, USA}

\author{Xuntao Wu}
\affiliation{Pritzker School of Molecular Engineering, University of Chicago, Chicago IL 60637, USA}

\author{Ming-Han Chou}
\altaffiliation[Present address: ]{AWS Center for Quantum Computing, Pasadena, CA 91125, USA}
\affiliation{Department of Physics, University of Chicago, Chicago IL 60637, USA}

\author{Gustav Andersson}
\affiliation{Pritzker School of Molecular Engineering, University of Chicago, Chicago IL 60637, USA}

\author{Christopher R. Conner}
\affiliation{Pritzker School of Molecular Engineering, University of Chicago, Chicago IL 60637, USA}

\author{Yash J. Joshi}
\affiliation{Pritzker School of Molecular Engineering, University of Chicago, Chicago IL 60637, USA}

\author{Shiheng Li}
\affiliation{Department of Physics, University of Chicago, Chicago IL 60637, USA}

\author{Jacob M. Miller}
\affiliation{Department of Physics, University of Chicago, Chicago IL 60637, USA}

\author{Rhys G. Povey}
\altaffiliation[Present address: ]{Institute for Quantum Optics and Quantum Information, Austrian Academy of Sciences, A-1090 Vienna, Austria}
\affiliation{Department of Physics, University of Chicago, Chicago IL 60637, USA}

\author{Hong Qiao}
\affiliation{Pritzker School of Molecular Engineering, University of Chicago, Chicago IL 60637, USA}

\author{Eric Chitambar}
\affiliation{Department of Electrical and Computer Engineering, University of Illinois at Urbana-Champaign, Urbana IL 61801, USA}

\author{Andrew N. Cleland}
\email{anc@uchicago.edu}
\affiliation{Pritzker School of Molecular Engineering, University of Chicago, Chicago IL 60637, USA}
\affiliation{Center for Molecular Engineering and Material Science Division, Argonne National Laboratory, Lemont IL 60439, USA}

\date{\today}

\maketitle
\setcounter{figure}{0}
\setcounter{equation}{0}
\setcounter{table}{0}
\renewcommand{\thefigure}{S\arabic{figure}}
\renewcommand{\thetable}{S\arabic{table}}
\renewcommand{\theequation}{S\arabic{equation}}
\clearpage
\section{Fabrication}
\begin{figure}[b]
    \centering
    \includegraphics{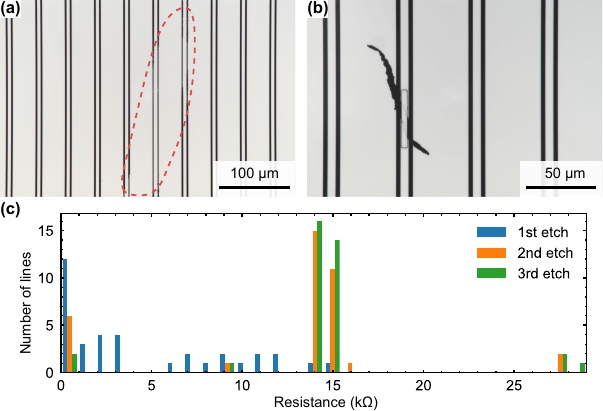}
    \caption{Fabrication of the long transmission lines. (a) Short indicated by the red dashed line fixed by a second etch. (b) Open fixed by a bandage layer. (c) Measured resistance of 36 1.3-meter-long lines on a 4-inch wafer after the first, second, and third etch.}
    \label{fig: longline}
\end{figure}
The sample fabrication follows a similar process to that described in Ref.~\cite{Grebel2024}. Circuits are fabricated on 100 mm diameter sapphire wafers. The aluminum base layer is first deposited by electron beam evaporation, and the circuit pattern is defined by reactive plasma etching through a photoresist stencil. The long coplanar waveguides (CPW) have an $8~\mathrm{\mu m}$ wide center conductor and a $4~\mathrm{\mu m}$ center conductor-ground gap, with a characteristic impedance of $50~\mathrm{\Omega}$. The 1.3-m-long lines are effectively shorted to ground on either end via gmon couplers. We include a test pad at the center point of each of these transmission lines. By measuring the resistance from the test pad to the ground, we can determine whether there is an unwanted intermediate open or short on the transmission line, due to fabrication flaws.

If a line is shorted to the ground before the gmon coupler at the end, caused e.g. by an incomplete etch of the base layer,  we locate the flaw and repair the defect by re-patterning a photoresist mask on the offending section of the line and performing a second etch; for the repair we reduce the CPW gap from $4~\mathrm{\mu m}$ to $2~\mathrm{\mu m}$ to reduce the impact of misalignments in the photolithography. In Fig.~\ref{fig: longline}(a), we show an example of such a repair. 

If a line is measured to have double the expected resistance, the center conductor has a break, due to a lithography flaw. Again, we locate the open using a microscope, then pattern an Al bandage layer to repair the break, as shown in Fig.~\ref{fig: longline}(b). 

In Fig.~\ref{fig: longline}(c), we display the measured resistances of 36 1.3-meter-long lines on one wafer, after the first (initial) etch, followed by a second and third repair etch. We can see that most of the shorts can be fixed with a second etch. The figure also shows a small number of lines with opens, that must be repaired as above.

We next deposit a lift-off-patterned $\mathrm{SiO_2}$ layer followed by a lift-off-patterned Al bandage layer for forming airbridges over the coplanar waveguide lines, and re-measure all the long transmission line resistances to verify there are no defects caused by the airbridges. The connections between the testpads and long transmission lines are etched away. We then pattern, deposit, and lift off the qubit Josephson junctions using a Dolan bridge process \cite{Dolan1977}. We singulate the wafers to $20\mathrm{mm}\times 20 \mathrm{mm}$ dies and remove the $\mathrm{SiO_2}$ airbridge support layer with an HF vapor etcher.

\section{Measurement setup}
\begin{figure}[t]
    \centering
    \includegraphics{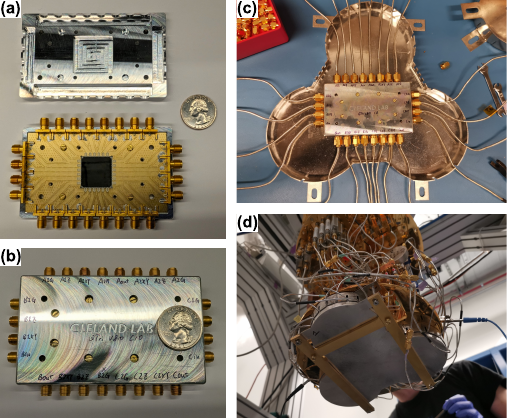}
    \caption{Sample packaging. (a-b) The sample is wire-bonded to a connectorized printed circuit board, which is mounted in an aluminum enclosure. (c-d) The aluminum enclosure is placed in a high magnetic permeability shield and mounted to the mixing chamber of a dilution refrigerator.}
    \label{fig: packaging}
\end{figure}
\begin{figure}[t]
    \centering
    \includegraphics{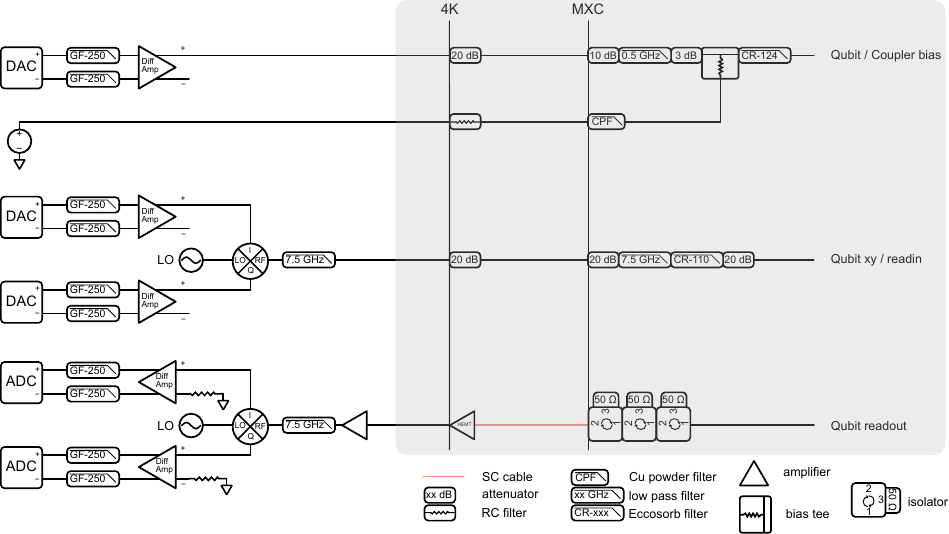}
    \caption{Measurement setup and wiring diagram.}
    \label{fig: wiring}
\end{figure}
The sample is wire-bonded to a connectorized printed circuit board (PCB) with Al wires and placed in an Al enclosure for measurement. The packaged sample is shown in Fig.~\ref{fig: packaging}(b). The enclosure has its lowest microwave resonance at about $7.5~\mathrm{GHz}$. The entire package is placed in a high-permeability magnetic shield, mounted on the mixing chamber stage of a dilution refrigerator, and cooled to $10~\mathrm{mK}$ prior to measurement. The measurement setup and wiring diagram are shown in Fig.~\ref{fig: wiring}. All the qubit $xy$ drive signals share a local oscillator (LO), while all the read-in and read-out channels share a separate LO. We place custom-made Eccosorb filters with NbTi center conductors on all the low-frequency flux lines \cite{Yan2024}.

\section{Sample characterization}
\begin{table*}[t]
  \caption{\label{tab:tableS1}Qubit characterization: qubit readout frequency $\omega_r$, qubit operating frequency $\omega_q$, qubit anharmonicity $\alpha$, readout dispersive shift $\chi_{ge}$, readout length $\tau_{rr}$, readout assigned probability for $|g\rangle$ ($|e\rangle$) state $F_g$ ($F_e$), single-qubit gate fidelity $F_{1Q}$, qubit-qubit coupling strength $g_{qq}$, qubit-qubit $ZZ$ coupling strength $\chi_{qq}$ and $\mathrm{CZ}$ gate fidelity $F_{CZ}$. For $T_{2,\mathrm{Ramsey}}$, we use the time evolution data before $300~\mathrm{ns}$ to do the fitting.}
  \begin{ruledtabular}
  \begin{tabular}{lcccccc}
  Qubit & $A_1$ & $A_2$ & $B_1$ & $B_2$ & $C_1$ & $C_2$\\\hline
  $T_1$ & $\mathrm{13~\mu s}$ & $\mathrm{7~\mu s}$ & $\mathrm{15~\mu s}$ & $\mathrm{12~\mu s}$ & $\mathrm{13~\mu s}$ & $\mathrm{10~\mu s}$\\
  $T_{2,\mathrm{Ramsey}}$& $\mathrm{170~ns}$& $\mathrm{2.6~\mu s}$& $\mathrm{1.8~\mu s}$& $\mathrm{1.8~\mu s}$& $\mathrm{1.7~\mu s}$& $\mathrm{4.5~\mu s}$\\
  $\omega_r/2\pi$ & $\mathrm{6.055~GHz}$ & $\mathrm{6.008~GHz}$ & $\mathrm{5.978~GHz}$ & $\mathrm{6.045~GHz}$ & $\mathrm{5.976~GHz}$ & $\mathrm{6.049~GHz}$\\
  $\omega_q/2\pi$ & $\mathrm{5.147~GHz}$ & $\mathrm{4.824~GHz}$ & $\mathrm{4.799~GHz}$ & $\mathrm{5.119~GHz}$ & $\mathrm{4.717~GHz}$ & $\mathrm{5.111~GHz}$\\
  $\alpha/2\pi$ & $\mathrm{-158~MHz}$ & $\mathrm{-167~MHz}$ & $\mathrm{-165~MHz}$ & $\mathrm{-166~MHz}$ & $\mathrm{-170~MHz}$ & $\mathrm{-171~MHz}$ \\
  $\chi_{ge}/2\pi$ & $\mathrm{-2.1~MHz}$ & $\mathrm{-2.0~MHz}$ & $\mathrm{-4.2~MHz}$ & $\mathrm{-3.0~MHz}$ & $\mathrm{-3.6~MHz}$ & $\mathrm{-2.6~MHz}$ \\
  $\tau_{rr}/2\pi$ & $\mathrm{820~ns}$ & $\mathrm{800~ns}$ & $\mathrm{720~ns}$ & $\mathrm{780~ns}$ & $\mathrm{560~ns}$ & $\mathrm{520~ns}$ \\
  $F_g$ & $98\%$ & $97\%$ & $99\%$ & $98\%$ & $99\%$ & $99\%$ \\
  $F_e$ & $93\%$ & $93\%$ & $95\%$ & $90\%$ & $94\%$ & $94\%$ \\
  $F_{1Q}$ & $99.5\%$ & $99.4\%$ & $99.7\%$ & $99.6\%$ & $99.6\%$ & $99.6\%$ \\\hline
  $g_{qq}/2\pi$&\multicolumn{2}{c}{$\mathrm{11.4~MHz}$}&\multicolumn{2}{c}{$\mathrm{11.4~MHz}$}&\multicolumn{2}{c}{$\mathrm{11.4~MHz}$}\\
  $\chi_{qq}/2\pi$&\multicolumn{2}{c}{$1.2~\mathrm{MHz}$}&\multicolumn{2}{c}{$1.3~\mathrm{MHz}$}&\multicolumn{2}{c}{$0.6~\mathrm{MHz}$}\\
  $F_{CZ}$&\multicolumn{2}{c}{N/A}&\multicolumn{2}{c}{$95.5\%$}&\multicolumn{2}{c}{$\mathrm{95.6\%}$}\\
  \end{tabular}
  \end{ruledtabular}
\end{table*}

Table~\ref{tab:tableS1} lists the qubit characterization data. We design all the readout resonator frequencies to be around $6~\mathrm{GHz}$ to avoid the enclosure mode at $7.5~\mathrm{GHz}$. The maximum qubit frequencies are above $6~\mathrm{GHz}$. The relatively short qubit dephasing times are due to the required large detuning ($>1~\mathrm{GHz}$) from the qubit maximum frequencies. Qubit $A_1$ has an unexpectedly short $T_{\phi}$, possibly caused by a noisy DC source, as we observed longer $T_{\phi}$ in a prior measurement, before the data in Table~\ref{tab:tableS1} was taken. All the qubit anharmonicities are around $-170~\mathrm{MHz}$. The slightly reduced anharmonicity, relative to our typical value of $-220~\mathrm{MHz}$, comes from the additional linear inductance in the gmon couplers. All single qubit rotations are performed by $30~\mathrm{ns}$ pulses with a cosine envelope and DRAG correction \cite{Motzoi2009}. We benchmark all the single-qubit gates with randomized benchmarking \cite{Knill2008} and get an average gate fidelity of around $99.6\%$. 

When using two adjacent qubits, we detune the neighboring qubit away from the target qubit by around $300~\mathrm{MHz}$. The $\mathrm{iSWAP}$ gates are realized by bringing two neighboring qubits on resonance, while the $\mathrm{CZ}$ gates are realized by bringing the two-qubit state $|gf\rangle$ on resonance with $|ee\rangle$. The coupling strength between two neighboring qubits $g_{qq}$ is measured by the duration of the $\mathrm{iSWAP}$ gate between them, which gives an $\mathrm{iSWAP}$ gate duration of $22~\mathrm{ns}$ and a $\mathrm{CZ}$ gate duration of $31~\mathrm{ns}$. The $ZZ$ interaction strength $\zeta_{qq}$ is measured by probing the frequency shift of one qubit depending on the state of the other qubit, measured by a Ramsey-type experiment. We benchmark the $\mathrm{CZ}$ gates by cross-entropy benchmarking \cite{Boixo2018} and get an average gate fidelity of around $95.6\%$. We did not benchmark the $\mathrm{CZ}$ gate between qubits $A_1$ and $A_2$, as this gate is not used in the experiments described here, due to the short $T_\phi$ of the qubit $A_1$.

\begin{table*}[t]
  \caption{\label{tab:tableS2}Communication channel characterization: communication channel frequency $\omega_c$, communication channel $T_1$ and $T_2$, transfer efficiency $\eta_t$ and generated Bell state fidelity $\mathcal{F}_{\mathrm{Bell}}$.}
  \begin{ruledtabular}
  \begin{tabular}{lccc}
   & $A_2-C_1$ & $C_2-B_2$ & $B_1-A_1$\\\hline
   $\omega_c/2\pi$ &$4.743~\mathrm{GHz}$ & $5.135~\mathrm{GHz}$& $4.905~\mathrm{GHz}$\\
  $T_1$ &$1.2~\mathrm{\mu s}$ & $1.0~\mathrm{\mu s}$& $1.0~\mathrm{\mu s}$\\
  $T_2$ &$2.3~\mathrm{\mu s}$ & $2.0~\mathrm{\mu s}$& $2.0~\mathrm{\mu s}$\\
  $\eta_t$ &$(88.7\pm 0.8)\%$ & $(89.0\pm 1.6)\%$& $(83.6\pm 3.0)\%$\\
  $\mathcal{F}_{\mathrm{Bell}}$ &$(95.1\pm 0.8)\%$ & $(95.3\pm 1.0)\%$& $(86.2\pm 0.5)\%$\\
\end{tabular}
  \end{ruledtabular}
\end{table*}

We benchmark the 1.3-m-long communication channels using the qubits as discussed in the main text. We swap an excitation from a qubit to the channel, wait, then swap back, allowing us to extract the channel relaxation times $T_1$ and dephasing times $T_2$. The measured results are listed in Table~\ref{tab:tableS2}. All the measured communication modes have $T_1$ around $1~\mathrm{\mu s}$, corresponding to a quality factor $Q$ of $\sim 3\times 10^4$, lower than the long CPW lines reported in Refs.~\cite{Zhong2019, Grebel2024}. The lower coherence time here might be due to a different circuit layout and the fabrication process.

We measure the qubit-line-qubit transfer efficiency $\eta_t$ by transferring a qubit excitation from one node to another. Bell states are generated by the half-transfer process ST/2 described in the main text. The generated Bell states are measured with state tomography, where we apply single qubit gates from the gate set $\{I, X/2, Y/2\}$ to each qubit after state preparation, and perform simultaneous readout of all qubits to get the probabilities on different axes of the Bloch sphere, and correct for readout errors \cite{Bialczak2010}. Assuming the measured state probabilities are $\mathbf{\widetilde{P}} = [\widetilde{P}_g,\widetilde{P}_e]^T = [1-\widetilde{P}_e,\widetilde{P}_e]^T$, the corrected state probabilities are $\mathbf{P}=[P_g,P_e]= \mathbf{F^{-1}}\mathbf{\widetilde{P}}$, where the readout correction matrix $\mathbf{F}$ is defined as 
\begin{equation}
    \mathbf{F}=\begin{bmatrix}
        F_g&1-F_e\\1-F_g&F_e
    \end{bmatrix}.
\end{equation}
For multiple qubits, we tensor the individual readout correction $\mathbf{F}$-matrices. We use convex optimization to reconstruct the density matrices to constrain them to be physical, i.e. Hermitian, unit trace, and positive semi-definite. The uncertainties of the measured $\eta_t$ and $\mathcal{F}_{\mathrm{Bell}}$ are determined by the standard deviation of 20 repeated measurements. The lower state transfer efficiency $\eta_t$ and Bell state fidelity $\mathcal{F}_{\mathrm{Bell}}$ between $B_1$ and $A_1$ comes from the short dephasing time of $A_1$ (see Table~\ref{tab:tableS1}).

We use the following Hamiltonian to model the state transfer and Bell state generation process,
\begin{equation}
\begin{split}
    H/\hbar = \omega_1 \sigma_1^\dagger \sigma_1 + \omega_2 \sigma_2^\dagger \sigma_2 + \sum_{m=1}^M\Big(\omega_c + \big(m-\frac{M+1}{2}\big)\omega_{\mathrm{FSR}}\Big) c_m^\dagger c_m\\
    + \sum_{m=1}^M g_1 (\sigma_1 c_m^\dagger +\sigma_1^\dagger c_m) + \sum_{m=1}^M (-1)^m g_2 (\sigma_2 c_m^\dagger + \sigma_2^\dagger c_m),
\end{split}
\end{equation}
where $\sigma_i$ and $\sigma_i^\dagger$ are the annihilation and creation operators for qubit $i$ at each node, $c_m$ and $c_m^\dagger$ are the annihilation and creation operators for the $m^{\rm th}$ communication mode, $M=5$ is the number of standing modes considered in the simulation, $\omega_{\mathrm{FSR}} = 50~\mathrm{MHz}$ is the free spectral range of the channel modes, and $g_i$ is the coupling strength between qubit $i$ and the $m^{\rm th}$ mode. We first excite qubit 1 and turn on the coupling $g_1/2\pi = 2.5~\mathrm{MHz}$ for a period $\tau_1$, then turn off $g_1$ and turn on $g_2/2\pi = 3.1~\mathrm{MHz}$ for a period $\tau_2=\pi/2g_2 =80.6~\mathrm{ns}$. When $\tau_1=\pi/2g_1 = 100~\mathrm{ns}$, we completely transfer the state from qubit $1$ to qubit $2$. When $\tau_2=\pi/4g_1=50 ~\mathrm{ns}$, a half-transfer, we generate a Bell state between two qubits. We use the measured qubit and channel $T_1$ and $T_2$ in all simulations. For the $A_2-C_2$ qubit pair, the simulated $\eta_t=89.8\%$ and $\mathcal{F}_{\mathrm{Bell}}=94.3\%$. For $C_2-B_2$, the simulated $\eta_t=89.3\%$ and $\mathcal{F}_{\mathrm{Bell}}=95.0\%$. For $B_1-A_1$, the simulated $\eta_t=80.2\%$ and $\mathcal{F}_{\mathrm{Bell}}=83.5\%$. The dominant loss comes from qubit dephasing and channel loss. 

\section{Entanglement swapping}
\begin{figure}[tb]
    \centering
    \includegraphics{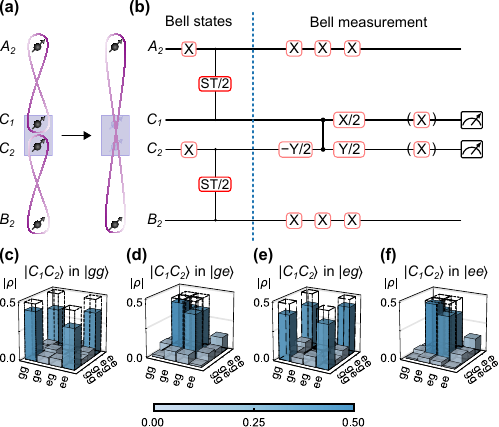}
    \caption{Entanglement swapping. (a) Schematic, (b) pulse sequence, and (c-f) states of $|A_2B_2\rangle$ after the swapping depending on the measurement results on $|C_1C_2\rangle$.}
    \label{fig: entanglement swapping}
\end{figure}

We can realize entanglement swapping \cite{Zukowski1993} using the scheme shown in Fig.~\ref{fig: entanglement swapping}(a). In entanglement swapping, we first prepare two Bell states in $A_2C_1$ and $C_2B_2$, and then do a Bell state measurement on $C_1C_2$, which maps $|C_1C_2\rangle$ to one of the Bell states. The resulting state $|A_2B_2\rangle$ will then also be a Bell state. Thus, we can entangle qubits $A_2$ and $B_2$ without using a direct communication channel between these qubits. 

In entanglement swapping, we need to perform a Bell state measurement, which projects the two-qubit states to Bell states. For superconducting qubits, the dispersive readout measures the qubits along the $z$-axis. We then need to transform the Bell states to product states and perform a readout, e.g. $|\psi^-\rangle\to |gg\rangle$. The pulse sequence for entanglement swapping is shown in Fig.~\ref{fig: entanglement swapping}(b). To better visualize how the circuit works, we write down the states after each step. The initial Bell state we generate between each pair of nodes is $|\psi^-\rangle=(|eg\rangle-|ge\rangle)/\sqrt{2}$; written in the four-qubit basis $|C_1C_2A_2B_2\rangle$, the state is 
\begin{equation}
    \frac{1}{2}(|ggee\rangle - |geeg\rangle - |egge\rangle + |eegg\rangle).
\end{equation}
The $-Y/2$, $Y/2$ and $\mathrm{CZ}$ gates together form a $\mathrm{CNOT}$ gate with $C_1$ as the control qubit and $C_2$ as the target qubit. After this $\mathrm{CNOT}$ gate, the state becomes
\begin{equation}
    \frac{1}{2}(|ggee\rangle - |geeg\rangle - |eege\rangle + |eggg\rangle).
\end{equation}
The $X/2$ gate transforms $|g\rangle$ to $(|g\rangle-i|e\rangle)/\sqrt{2}$ and $|e\rangle$ to $(-i|g\rangle+|e\rangle)/\sqrt{2}$, which results in the state
\begin{equation}
\begin{split}
    &\frac{1}{2\sqrt{2}}(|ggee\rangle -i|egee\rangle - |geeg\rangle + i|eeeg\rangle + i|gege\rangle- |eege\rangle -i|gggg\rangle + |eggg\rangle)\\
    =&\frac{1}{2\sqrt{2}}\big(|gg\rangle(|ee\rangle-i|gg\rangle)-|ge\rangle(|eg\rangle-i|ge\rangle)+|eg\rangle(|gg\rangle-i|ee\rangle) -|ee\rangle(|ge\rangle-i|eg\rangle) \big).
\end{split}
\end{equation}
We also apply three dynamical decoupling (DD) $X$ gates on $A_2$ and $B_2$ respectively to reduce dephasing effects. After these $X$ gates, we get the output states of $|A_2B_2\rangle=|\psi_{mn}\rangle$ when $|C_1C_2\rangle$ is measured in $|mn\rangle$,
\begin{align}
    |\psi_{gg}\rangle = \frac{1}{\sqrt{2}}(|gg\rangle-i|ee\rangle),\\
    |\psi_{ge}\rangle = \frac{1}{\sqrt{2}}(|ge\rangle-i|eg\rangle),\\
    |\psi_{eg}\rangle = \frac{1}{\sqrt{2}}(|gg\rangle+i|ee\rangle),\\
    |\psi_{ee}\rangle = \frac{1}{\sqrt{2}}(|ge\rangle+i|eg\rangle).
\end{align}
These are all maximally entangled states with a phase difference from the Bell states. If we replace the $X/2$ gate with a $Y/2$ or a Hadamard $H$ gate, all the output states will be Bell states. We perform state tomography on all four possible output states and find state fidelities of $(77.9\pm 1.6)\%$, $(81.5\pm 1.5)\%$, $(80.2\pm 1.2)\%$, $(81.8\pm 1.3)\%$, with the density matrices plotted in Fig.~\ref{fig: entanglement swapping}(c-g). To further reduce the measurement errors, we apply optional $X$ gates on $C_1$ and $C_2$ to map $|C_1C_2\rangle$ to $|ee\rangle$ when performing tomography on the target output states. The performance is limited by gate errors, measurement errors, and dephasing in qubits $A_2$ and $B_2$ during local operations on qubits $C_1$ and $C_2$.

\section{GHZ state generation}
\begin{figure}[t]
    \centering
    \includegraphics{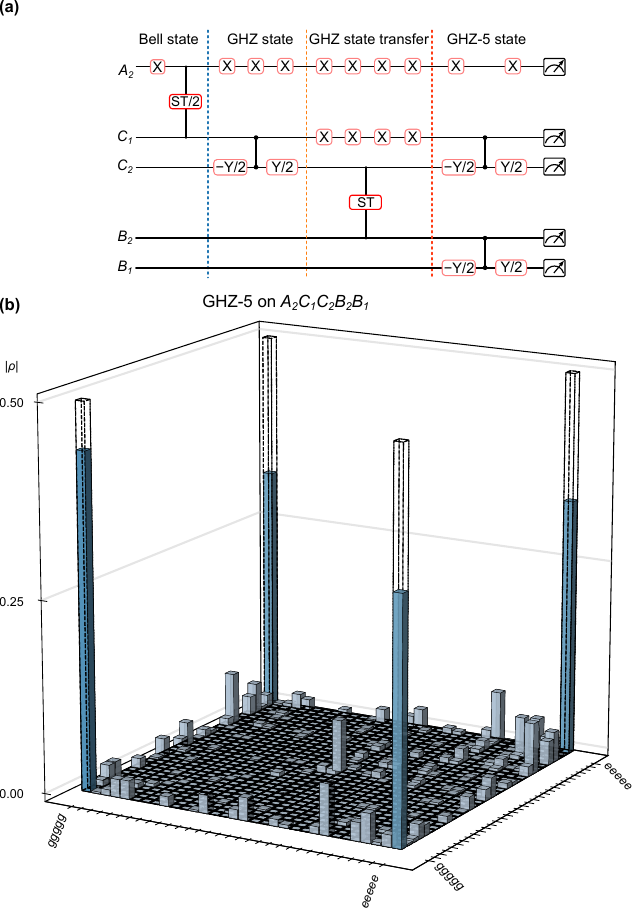}
    \caption{Generation of GHZ states. (a) Pulse sequence to generate a GHZ-5 state on $A_2C_1C_2B_2B_1$. (b) Density matrix of GHZ-5 with a fidelity of $70.3\pm 1.1\%$.}
    \label{fig: GHZ}
\end{figure}

In the main text, we discuss the GHZ state generation on $A_2C_1B_2$. Here we expand the GHZ-3 state to a GHZ-5 state on $A_2C_1C_2B_2B_1$ using the circuit shown in Fig.~\ref{fig: GHZ}(a). The measured density matrix with fidelity of $(70.3\pm 1.1)\%$ is shown in Fig.~\ref{fig: GHZ}(b).

\section{Privacy bound for quantum secret sharing}
\begin{figure}[t]
    \centering
    \includegraphics{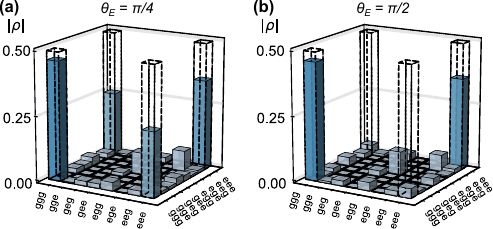}
    \caption{The measured density matrices for $A_2C_1B_2$ when (a) $\theta_E=\pi/4$ and (b) $\pi/2$.}
    \label{fig: eavesdropping_tomo}
\end{figure}
The measured density matrices on $A_2C_1B_2$ when $\theta_E=\pi/4$ and $\pi/2$ are shown in Fig.~\ref{fig: eavesdropping_tomo}.

Given the correspondence between prepare-and-measure protocols and entanglement-based protocols, in below we first consider the prepare-and-measure protocol and derive the privacy bound, which can then be reformulated in the entanglement-based scenario.

\subsection{Prepare-and-measure protocol}
The sender Alice prepares an ensemble of quantum states $\{\rho_{Ai}\}$, and she randomly picks a state from the ensemble according to probability distribution $\{p_i\}$ which will be sent over a quantum channel $\mathcal{E}$ to Bob. Bob performs measurements to extract information from the received quantum states $\rho_{Bi}=\mathcal{E}(\rho_{Ai})$. The sent ensemble is $\rho_A=\sum_ip_i\rho_{Ai}$, and the received ensemble is $\rho_B=\sum_ip_i\rho_{Bi}$. Then the receiver will obtain $I(A:B)$ amount of classical information that the sender wishes to convey. Meanwhile, the eavesdropper may also have access to the sent information by tapping the channel, behaving as (part of) the environment that implements the observed channel $\mathcal{E}$. The eavesdropper will obtain $I(A:E)$ amount of classical information that the sender wishes to convey.

When the receiver obtains more information than the eavesdropper, the sender and the receiver can in principle use the excess information to distill secret keys between them. Therefore, we may define the privacy of the communication channel as~\cite{Schumacher1998}
\begin{align}
    P = I(A:B) - I(A:E).
\end{align}
The sender and the receiver have no control over the eavesdropper, so we may assume the eavesdropper has the greatest possible power. Correspondingly, we may then define the ``guaranteed'' privacy $p_G = \inf_\mathrm{E}P$, where the infimum is taken over all possible strategies of the eavesdropper to maximize $I(A:E)$. According to the Holevo bound, the information that the eavesdropper can obtain is upper bounded by the Holevo quantity
\begin{equation}
    I(A:E)\leq S(\rho_E)-\sum_ip_iS(\rho_{Ei})=\chi_E.
\end{equation}
Therefore, we have the lower bound of the guaranteed privacy
\begin{align}
    p_G \geq I(A:B) - \chi_E.
\end{align}

The sender and the receiver may also want to increase the guaranteed privacy $p_G$, i.e. we are interested in the optimal guaranteed privacy $\mathcal{P}=\sup_{AB}p_G$, where the supremum is taken over all possible of strategies of sending and receiving between Alice and Bob by using the channel $\mathcal{E}$. Again, the Holevo bound tells us that 
\begin{equation}
    I(A:B)\leq S(\rho_B)-\sum_ip_iS(\rho_{Bi})=\chi_B.
\end{equation}
According to Holevo-Schumacher-Westmoreland (HSW) theorem~\cite{Holevo1998,Schumacher1997}, the (single-copy) Holevo bound on mutual information $\chi_B$ is achievable asymptotically (through sending $n\gg 1$ copies). Therefore, the optimal guaranteed privacy should statisfy
\begin{align}
    \mathcal{P} \geq \chi_B - \chi_E.
\end{align}

Note that $\mathcal{P}$ also depends on the state ensemble to be sent, in addition to the transmission channel. We may need to further optimize over possible state ensembles to get the optimal guaranteed privacy. Nevertheless, we can always consider a specific state ensemble to derive an explicit lower bound. Consider that Alice wants to send an ensemble of pure states $\{(p_i,|\psi_i\rangle)\}$, where $p_i$ is the probability for choosing to send $|\psi_i\rangle$, and we denote $\rho=\sum_ip_i|\psi_i\rangle\langle\psi_i|$. We also assume that the entire channel $\mathcal{E}$ is due to Eve's tapping as 
\begin{align}
    \mathcal{E}(O_B) = \mathrm{Tr}_E\left[U_{BE}(O_B\otimes|0\rangle\langle 0|_E)U_{BE}^\dagger\right],
\end{align}
where $|0\rangle_E$ is a certain initial state of the eavesdropper. Then the joint transmitted states between Bob and Eve is $|\Psi_i\rangle_{BE}=U_{BE}(|\psi_i\rangle_B|0\rangle_E)$, and the received states by Bob and Eve are
\begin{align}
    \rho_{Bi} = \mathrm{Tr}_E(|\Psi_i\rangle\langle\Psi_i|_{BE}),~\rho_{Ei} = \mathrm{Tr}_B(|\Psi_i\rangle\langle\Psi_i|_{BE}),
\end{align}
which satisfy $S(\rho_{Bi})=S(\rho_{Ei})$. We then have 
\begin{align}\label{eqn:opt_guarantee_privacy}
    \mathcal{P} \geq S(\mathcal{E}(\rho)) - S(\sum_ip_i\rho_{Ei}) = S(\mathcal{E}(\rho)) - S(\rho_E).
\end{align}

\subsection{Entanglement-based protocol}
We experimentally implemented the GHZ-based QSS protocol, which is not the same as prepare-and-measure protocols. However, it can be interpreted as a prepare-and-measure protocol, where the sender ``sends'' a quantum state to the receivers by measuring the subsystem that they hold depending on the measurement result. Although there are three parties in the QSS experiment, here we focus on external eavesdropping and thus the two receivers Bob and Charlie can be viewed as a whole. We can assume that the noise in the GHZ states all comes from the eavesdropping, when the sender distributes the GHZ state to the receivers, that is
\begin{align}
    \rho_{ABC} = \mathrm{Tr}_E\left[(I_A\otimes U_{BCE})(|\mathrm{GHZ}\rangle\langle\mathrm{GHZ}|_{ABC}\otimes|0\rangle\langle 0|_E)(I_A\otimes U_{BCE}^\dagger)\right].
\end{align}
Without loss of generality, we consider the case where Alice measures in the $x$ basis. The joint state between Alice, Bob, Charlie and Eve is
\begin{align}
    |\Psi\rangle_{ABCE} =& (I_A\otimes U_{BCE})(|\mathrm{GHZ}\rangle_{ABC}\otimes|0\rangle_E)\nonumber\\
    =& \frac{1}{\sqrt{2}}|x+\rangle_A\otimes U_{BCE}\left[\frac{1}{\sqrt{2}}(|x+\rangle_B|x+\rangle_C + |x-\rangle_B|x-\rangle_C)\otimes|0\rangle_E\right]\nonumber\\
    & + \frac{1}{\sqrt{2}}|x-\rangle_A\otimes U_{BCE}\left[\frac{1}{\sqrt{2}}(|x+\rangle_B|x-\rangle_C + |x-\rangle_B|x+\rangle_C)\otimes|0\rangle_E\right]\nonumber\\
    =& \frac{1}{\sqrt{2}}\left(|x+\rangle_A\otimes |\Psi_{x+}\rangle_{BCE} + |x-\rangle_A\otimes |\Psi_{x-}\rangle_{BCE}\right).
\end{align}
We then have the following interpretation. There is 1/2 probability for Alice to ``send'' the $(|x+\rangle_B|x+\rangle_C + |x-\rangle_B|x-\rangle_C)/\sqrt{2}$ state to Bob and Charlie, and 1/2 probability to ``send'' the $(|x+\rangle_B|x-\rangle_C + |x-\rangle_B|x+\rangle_C)/\sqrt{2}$ state. The states are ``sent'' through the effective channel 
\begin{align}
    \mathcal{E}(O_{BC}) = \mathrm{Tr}_E\left[U_{BCE}(O_{BC}\otimes|0\rangle\langle 0|_E)U_{BCE}^\dagger\right].
\end{align}
The received ensemble by Bob and Charlie is
\begin{align}
    p_{x+}\rho_{BC,x+} + p_{x-}\rho_{BC,x-} =& \frac{1}{2}\mathrm{Tr}_E\left(|\Psi_{x+}\rangle\langle|\Psi_{x+}|_{BCE}\right) + \frac{1}{2}\mathrm{Tr}_E\left(|\Psi_{x-}\rangle\langle|\Psi_{x-}|_{BCE}\right)\nonumber\\
    =& \mathrm{Tr}_A\left[\mathrm{Tr}_E\left(|\Psi\rangle\langle\Psi|_{ABCE}\right)\right]\nonumber\\
    =& \mathrm{Tr}_A\left(\rho_{ABC}\right).
\end{align}
Similarly the received ensemble by Eve is
\begin{align}
    p_{x+}\rho_{E,x+} + p_{x-}\rho_{E,x-} =& \frac{1}{2}\mathrm{Tr}_{BC}\left(|\Psi_{x+}\rangle\langle|\Psi_{x+}|_{BCE}\right) + \frac{1}{2}\mathrm{Tr}_{BC}\left(|\Psi_{x-}\rangle\langle|\Psi_{x-}|_{BCE}\right)\nonumber\\
    =& \mathrm{Tr}_{BC}\left[\mathrm{Tr}_A\left(|\Psi\rangle\langle\Psi|_{ABCE}\right)\right]\nonumber\\
    =& \mathrm{Tr}_{ABC}\left(|\Psi\rangle\langle\Psi|_{ABCE}\right)
\end{align}
Then according to Eqn~\ref{eqn:opt_guarantee_privacy}, we have that
\begin{align}
    \mathcal{P} \geq S(\mathrm{Tr}_A(\rho_{ABC})) - S(\mathrm{Tr}_{ABC}(|\Psi\rangle\langle\Psi|_{ABCE})) = S(\rho_{BC}) - S(\rho_{ABC}).
\end{align}

\subsection{Devetak-Winter distillable key rate}
We can also consider the one-way distillable key rate derived by Devetak and Winter~\cite{Devetak2004,Devetak2005}. Consider the classical-quantum (cq) state
\begin{align}
    \rho_{ABE} = \sum_xP(x)|x\rangle\langle x|_A\otimes\rho_{BE,x},
\end{align}
where $|x\rangle$ are orthonormal to each other, and $P(x)$ defines the probability distribution of a random variable $X$. It is then proved~\cite{Devetak2004,Devetak2005} that if we want to use one-way public communication to distill a secret key between A and B (in the QSS setup B corresponds to both Alice and Bob), given shared $\rho_{ABE}$ state, the distillable key rate $K$ satisfies
\begin{align}
    K \geq I(X:B) - I(X:E),
\end{align}
where $I(X:B)$ and $I(X:E)$ here denote the \textit{quantum mutual information} between A and B, and between A and E, respectively, while in the cq state subsystem A has become a classical random variable $X$. For bipartite states $\rho_{AB}$ the quantum mutual information is $I(A:B)=S(\rho_A)+S(\rho_B)-S(\rho_{AB})$, where $\rho_{A(B)}=\mathrm{Tr}_{B(A)}\rho_{AB}$. For the cq state considered above, if we assume the eavesdropper the greatest power so that $\rho_{BE,x}=|\psi_x\rangle\langle\psi_x|_{BE}$ is pure, we have
\begin{align}
    I(X:B) =& S\left(\sum_xP(x)|x\rangle\langle x|\right) + S\left(\sum_xP(x)\mathrm{Tr}_E|\psi_x\rangle\langle\psi_x|_{BE}\right)\nonumber\\
    &- S\left(\sum_xP(x)|x\rangle\langle x|_A\otimes\mathrm{Tr}_E|\psi_x\rangle\langle\psi_x|_{BE}\right)\nonumber\\
    =& -\sum_xP(x)\log P(x) + S(\rho_B) - S\left(\sum_xP(x)|x\rangle\langle x|_A\otimes\sum_i\lambda_{xi}|\phi_{xi}\rangle\langle\phi_{xi}|_B\right) \nonumber\\
    =& -\sum_xP(x)\log P(x) + S(\rho_B) + \sum_x\sum_i[P(x)\lambda_{xi}]\log [P(x)\lambda_{xi}] \nonumber\\
    =& -\sum_xP(x)\log P(x) + S(\rho_B) + \sum_x\sum_i[P(x)\lambda_{xi}]\log P(x)\nonumber\\
    &+ \sum_x\sum_i[P(x)\lambda_{xi}]\log\lambda_{xi} \nonumber\\
    =& -\sum_xP(x)\log P(x) + S(\rho_B) + \sum_xP(x)\left(\sum_i\lambda_{xi}\right)\log P(x)\nonumber\\
    &- \sum_xP(x)\left(-\sum_i\lambda_{xi}\log\lambda_{xi}\right) \nonumber\\
    =& S(\rho_B) - \sum_xP(x)S(\rho_{Bx}),
\end{align}
where $\rho_{Bx} = \mathrm{Tr}_E|\psi_x\rangle\langle\psi_x|_{BE}$, $\rho_B=\sum_xP(x)\rho_{Bx}$. We have also used the spectral decomposition $\rho_{Bx} = \sum_i\lambda_{xi}|\phi_{xi}\rangle\langle\phi_{xi}|_B$, so that $|x\rangle_A\otimes|\phi_{xi}\rangle_B$ are orthonormal. We can see that the above is exactly the Holevo quantity $\chi_B$. The same applies to $I(X:E)$ under the assumption that $\rho_{BE,x}=|\psi_x\rangle\langle\psi_x|_{BE}$ is pure. Therefore, the cq state can be interpreted as the result of measuring the shared entangled state, such as the GHZ state in the QSS setup. Then, we have
\begin{align}
    K \geq S(\rho_B) - S(\rho_E),
\end{align}
which is the same as the previously derived optimal guaranteed privacy bound. Again recall that in the QSS setup the subsystem B includes both Bob and Charlie.

On the other hand, we may also choose to use other experimental data to calculate the lower bound of $K$. For instance, considering that $I(X:E)\leq S(\rho_{ABC})$ we have
\begin{align}
    K \geq I(X:B) - S(\rho_{ABC}).
\end{align}
We can further make projective measurements in $x$ or $y$ basis of qubit A in the GHZ state, and perform state tomography of qubits B and C conditioned on different measurement results of A to obtain $\rho_{BCx}$, $x=0,1$. Then we can construct the cq state and calculate $I(X:B)$ explicitly.

\bibliographystyle{unsrt}
\bibliography{QSS_ref}